# Genesis of the James Webb Space Telescope architecture: The designers' story

**Pierre Y. Bely,[a]\*† , Garth D. Illingworth,[b]  Jonathan W. Arenberg,[c] Charles Atkinson,[c] Richard Burg,[d]  Mark Clampin,[d] Lee D. Feinberg,[d] Paul H. Geithner,[d] John C. Mather,[d] Michael T. Menzel,[d]  Max Nein,[e]\* Larry Petro,[a]\* David  C. Redding,[f] Bernard D. Seery,[d] H. Philip Stahl,[e]  Massimo Stiavelli,[a]  Hervey Stockman,[a]\* Scott P. Willoughby[e]**

a Space Telescope Science Institute, 3700 San Martin Drive, Baltimore, MD, 21218, USA

b UCO/Lick Observatory, University of California, Santa Cruz, CA 95064, USA

c Northrop Grumman, One Space Park Dr., Redondo Beach, CA, 90278, USA

d NASA Goddard Space Flight Center, 8800 Greenbelt Road, Greenbelt, MD 20771, USA

e NASA Marshall Spaceflight Center, Redstone Arsenal, Huntsville, AL 35812, USA

f NASA Jet Propulsion Laboratory, 4800 Oak Grove Drive, La Cañada Flintridge, CA, 91011, USA

\*Retired

Affiliations correspond to those valid at the time of primary involvement in the project.

## Abstract

The James Webb Space Telescope, launched in 2021, is an infrared observatory of novel design: deployable, with active optics, fully open to space for radiative cooling and orbiting the Lagrange point no. 2.  This article explains the rationale leading to this specific design and describes the various other architectures that were considered along the way: from a monolithic 10-meter telescope in geosynchronous orbit to a 6-meter one in High Earth Orbit, then a 16-meter observatory on the Moon, a 4- or 6-meter one in an elliptical heliocentric orbit, and a segmented 8-meter one passively cooled on 50 K at L2, which was finally descoped to 6.6 meters. It also addresses the optimization for scientific performance, the challenge of dealing with such an ultra-low operating temperature, cost issues, supporting technology, modifications made during final design and, finally, how the architecture performs on orbit.

†Pierre Y. Bely, email: pybely@gmail.com

## 1. Introduction

The James Webb Space Telescope (JWST), arguably one of the most powerful and complex devices yet devised to probe the Universe, is an unusual observatory.  Extremely large, deployable, not enclosed in a tube but fully open to space and located far from the Earth, it is unlike any other telescope observing in the same spectral range.  It bears no resemblance to Galileo's "eyeglass" which revolutionized our understanding of the world around us, nor to Hale's 100-inch telescope which transformed our view of the Universe, nor even to its modern predecessor, the Hubble Space Telescope (HST), which had thus far produced the deepest views of the Universe.

The description of JWST, the power of its scientific potential and its fabrication have already been the subject of several articles[1,2,3,4,5,6,7,8], but not much has been written about how it came to look as it does. What follows is the story of JWST's various incarnations and the genesis of its architecture as recounted by major actors in its conception and realization, drawing upon published documents as well as on personal notes and recollections.

In the context of space programs, "architecture" refers to the high-level organization, operating principles and design of a mission and, in the interest of conciseness, will be limited here to the observatory proper. Space support systems and science instruments will not be covered and details about the scientific and programmatic background of JWST can be found elsewhere[2,4].

## 2. A man, a vision

Although space agencies are generally in charge of the management and technical operation of space missions, their scientific operation is typically entrusted to a group of scientists, engineers and technicians at a university or research laboratory. In the case of the Hubble Space Telescope (HST), a specific institute was created for this purpose: the Space Telescope Science Institute (STScI). The Institute was established in 1981 and located on the Johns Hopkins University Campus in Baltimore, Maryland,[9,10] with Riccardo Giacconi as its first director (Fig. 1).

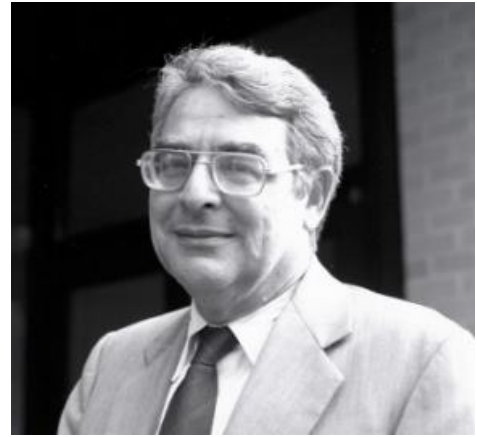

**Fig. 1.** Riccardo Giacconi at STScI in 1985. (Photo by P. Bely)

As it happened, Giacconi was not only a highly regarded scientist but also had experience in the design and construction of several important space satellites (the X-Ray satellites Uhuru, Einstein, and AXAF/Chandra). He was therefore acutely aware that, from inception to completion of any space-based astronomical observatory, much time elapses, typically 10 to 15 years. For HST it had taken over 20 years: Lyman Spitzer, the father of HST, had started lobbying for it back in the 1960s. So, in 1985, before HST was even launched, Giacconi was convinced that it was time to begin thinking about its successor. Indeed, since HST had a planned lifetime of only 14 years from its projected launch date of 1987, it was already rather late to start.

Why even build a successor? After all, HST was a fairly large telescope (2.4 m in diameter) with a spectral range and spatial resolution unmatched by ground telescopes and capable of seeing to the edge of the Universe. It could possibly be the only space telescope that would ever be needed. And if not, wouldn't it be better to wait for HST to show what it could do before trying to define a successor?

But the history of astronomy has shown that such thinking lacks foresight. At the dawn of the 20th century astronomers had a limited understanding of the Universe. Then, with the big jump in size to the 100-inch telescope on Mt Wilson, Edwin Hubble was able to prove that there were other worlds beyond our own galaxy. And in 1964 a giant antenna built for commercial communication allowed Penzias and Wilson to discover the cosmic background radiation by chance. In fact, as Martin Harwit has demonstrated[11], many important astronomical discoveries followed on the heels of major improvements in observational capabilities and were entirely unexpected.

The observational capability of an instrument using electromagnetic radiation to gather information about the Universe depends primarily on four factors: sensitivity, spatial resolution, spectral coverage and field of view. Increased sensitivity allows one to see objects that are fainter and is a function of the collecting area (i.e. the square of the diameter of the telescope for a circular aperture); increased spatial resolution allows one to see more detail in the observed objects and, again for a telescope with circular aperture, is proportional to its diameter; increased spectral coverage permits observation of objects via different wavelengths or types of radiation (e.g. radio, infrared, UV, X-rays) and helps analyze the physical phenomena at play; a greater field of view allows for the observation of large objects or the simultaneous observation of many small objects, as in surveys. While astronomical telescopes had sometimes been built for specific types of research, it seemed clear that a true successor to HST had to be a general-purpose observatory. And as such, there was no need to wait for HST to show its limitations: building a telescope with a larger diameter and wider spectral coverage would inevitably enable new science and result in major new discoveries.

### 3. First attempt at an HST successor

Although Giacconi was convinced that work on a successor to HST should commence promptly, STScI was only in charge of the scientific operation of the Hubble telescope, not staffed for instrumental design. Several astronomers there were well acquainted with the technical aspects of telescopes, in particular Deputy Director Garth Illingworth and Research Branch Chief Peter Stockman, and there was a substantial number of engineers, mostly software engineers, working on the development of the observation scheduling system. But STScI's Chief Engineer, Pierre Bely, did have experience in large optical telescope design. Having previously served as Project Engineer for the construction of the Canada-France-Hawaii Telescope, he had been hired by the Institute to monitor the technical aspects of HST. So, in April of 1985, Giacconi tasked Bely to come up with some ideas for a super HST. To increase its discovery space over HST, Giacconi, in concert with Illingworth and Stockman, stipulated that its spectral coverage should extend into the mid-infrared*, say 10 μm, albeit with moderate sensitivity. And since this was at a time when several 8–10-meter ground-based telescopes were already planned, the future space telescope should be in the same range, preferably a 10-meter instrument, in order to stay well ahead of competition from the ground (Fig. 2).

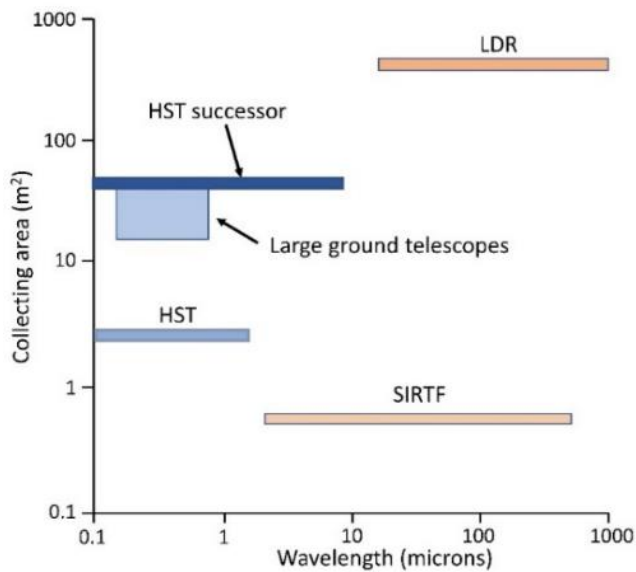

**Fig. 2.** Sensitivity (measured as collecting area) and spectral coverage of a 10-meter HST successor compared to that of planned space telescopes (Space InfraRed Telescope Facility, SIRTF, later renamed Spitzer, and the Large Deployable Reflector, LDR) and 8-to 10-meter ground telescopes. Not shown in this graph is a third axis, resolution, where space telescopes have a clear advantage over ground telescopes. (Bely/STScI)

Bely did not have to start from scratch. In the early 1980's, the Advanced Concepts Office at NASA's Marshall Space Flight Center (MSFC) had already begun investigating future potential astronomical space missions for long-term planning purposes. These included:

- VLST, the Very-Large-Space-Telescope for Optical/UV Astronomy[12], an 8-meter telescope, basically a scaled-up HST but with a segmented mirror, to be launched by the Space Shuttle, assembled by astronauts, and placed in a Low Earth Orbit (Fig. 3).

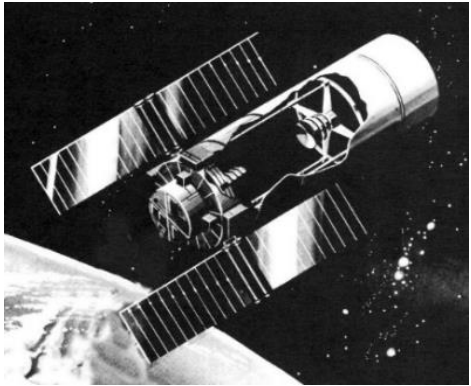

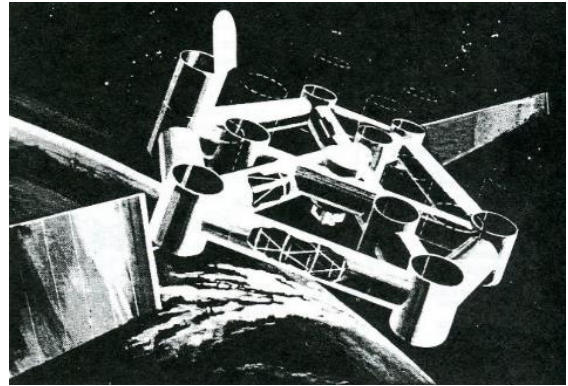

**Fig. 3.** Artist's view of the VLST.
(MSFC - 1980)

**Fig. 4.** Artist's view of the Golay-9 co-phased array.
(MSFC- 1982)

- A Golay-9 array, an array of nine 1.7-meter telescopes arranged in non-redundant configuration[13] (Fig. 4), or COSMIC (Coherent Optical System of Modular Imaging Collectors)[14], composed of four co-phased telescopes 1.8-meters in diameter aligned over a baseline of 15 meters. Imagery was obtained by rotation of the array and aperture synthesis. They were to be launched by the Space Shuttle, assembled by astronauts, and placed in Low Earth Orbit.

- The Large Deployable Reflector (LDR), a 20-meter diameter, far infrared and sub-millimetric telescope under study by the Jet Propulsion Laboratory[15,16] (Fig. 5). The LDR had a segmented mirror and was passively cooled to about 200 K by radiation to the sky. It was to be assembled by astronauts and stationed close to the International Space Station for resupply in cryogens.

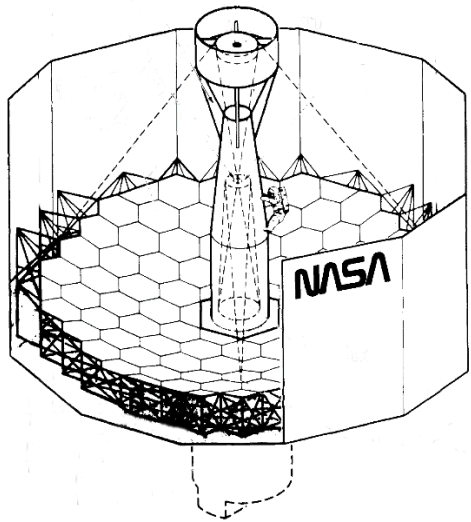

**Fig. 5.** Artist's view of the LDR. The astronaut climbing the central baffle gives the scale. (JPL- 1986)

- In parallel with these studies, NASA was funding research at the University of Arizona on the fabrication of large lightweight monolithic glass mirrors of very fast aperture[17,18]. Twenty years later this research would lead to the successful polishing of the f/1.14, 8.4-meter diameter mirrors for the Large Binocular Telescope (LBT)[19] (Fig. 6).

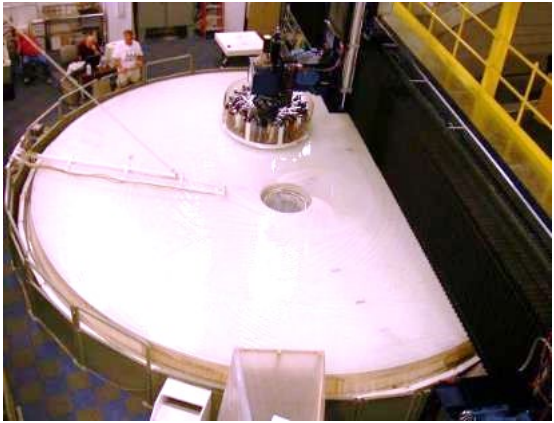

**Fig. 6.** Stressed lap polishing of the LBT's second mirror at the Steward Observatory Mirror Lab in 2002. (Steward Observatory)

With these concepts and studies in mind, the following conclusions were reached:

The VLST was a potential candidate for a successor to HST, but it would have to be cold for IR operation. However, its design and location in a low Earth orbit accessible to the Shuttle did not lend itself to cold, stable operation. The co-phased array, such as the Golay 9, had a potential fabrication and cost advantage, as it is easier to manufacture and launch several moderate-sized telescopes than a single large one. It also had the advantage of providing higher resolution for the same collecting area, since resolution increases with overall size. However, this approach is only beneficial for bright objects. As in HST, the forte of its successor needed to be the observation of very distant objects in the Universe, i.e., faint sources in the presence of a non-negligible background (celestial, self-emission, detector dark current). Diluted apertures are not good for that (Fig. 7). Image concentration could be improved by bringing the individual telescopes close together, but arrays still suffer from other ailments: very small field and poor throughput because of losses in the beam transfers. For faint sources the observations are background-limited and require high sensitivity. As such, a compact, single-surface aperture is essential.

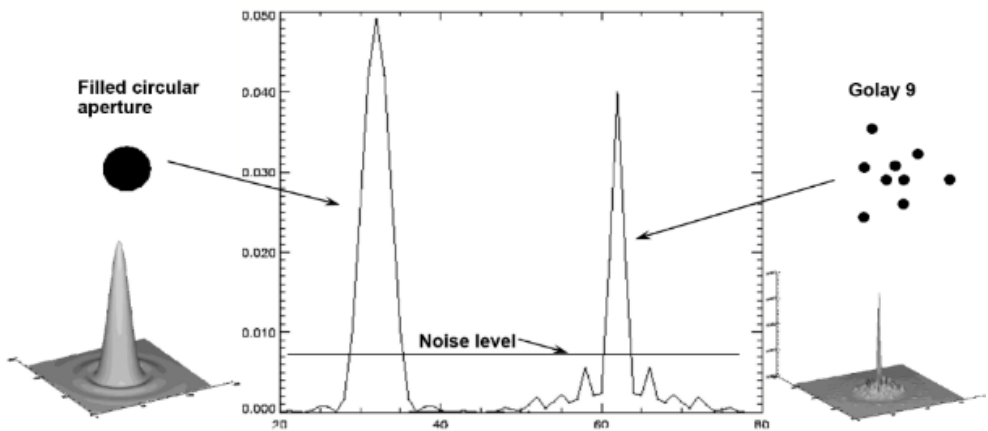

**Fig. 7.** The point spread function (PSF) of a filled circular aperture compared to that of a co-phased array (here Golay-9) with the same total collecting area, in the presence of significant background. The central peak of the Golay-9 PSF is narrower, providing a higher resolution, but much of the light is lost in background noise and cannot be recovered by image processing. (Bely/STScI)

Then what about the LDR? Diffraction limited at 30 μm and much too warm, it did not lend itself to a simple scaling-down. It would be better to start from scratch. But the detailed thermal study that was performed for LDR was of interest because it showed that the optics could be cooled to as low as 200 K by purely passive means, using a relatively short external shield to increase exposure to deep space and boosting the orbit altitude to 700 km in order to reduce heat input from the Earth.

Based on these considerations, Bely and François Roddier, an expert in adaptive optics and interferometric imaging from the National Optical Astronomy Observatories in Tucson, Arizona, developed the concept of a 10-meter telescope in space with the following rationale (Fig. 8):

1. Site the observatory in a geo-stationary orbit to minimize thermal radiation and straylight from Earth (with the side benefit of facilitating communications).

2. Take advantage of progress in the polishing of large mirrors by using a monolithic primary rather than a more complicated segmented one and reducing the telescope's length by using a fast aperture (f/1).

3. Use a minimal light shield to increase the observatory's solid angle view of deep space so as to maximize radiative cooling and allow the optics to reach a temperature on the order of 150 K, providing a foray into the mid-infrared.

4. Relax optics fabrication and structural stability tolerances by incorporating a minimum of active optics and by requiring that the optics be diffraction-limited only in the near infra-red. Use interferometry techniques as previously proposed by Roddier[20] to obtain diffraction-limited image quality in the UV and visible over a small field.

5. Launch the observatory fully assembled and tested using the USSR's heavy lift launcher Energia with an enlarged shroud – a possibility in view of the U.S.-Soviet rapprochement at the time.

Bely and Roddier presented their concept at the Aerospace Sciences meeting of the American Institute of Aeronautics and Astronautics Conference in January 1986[21], four years prior to the launch of HST. Bely presented a variant of it at two additional conferences that year[22,23], the last one with a primary mirror diameter reduced to 6 meters, so as to be compatible with planned US heavy lift launchers.

Work on the concept continued at a low level over the next few years in collaboration with John Bolton from NASA's Goddard Space Flight Center (GSFC), Phillip Tulkoff from Swales Aerospace, and Scott May from Johns Hopkins University concerning the passive cooling of the observatory, the structural and thermal aspects of the ultra light-weight, self-supported primary mirror[24,25] and the pointing of the observatory assisted by active optics[26].

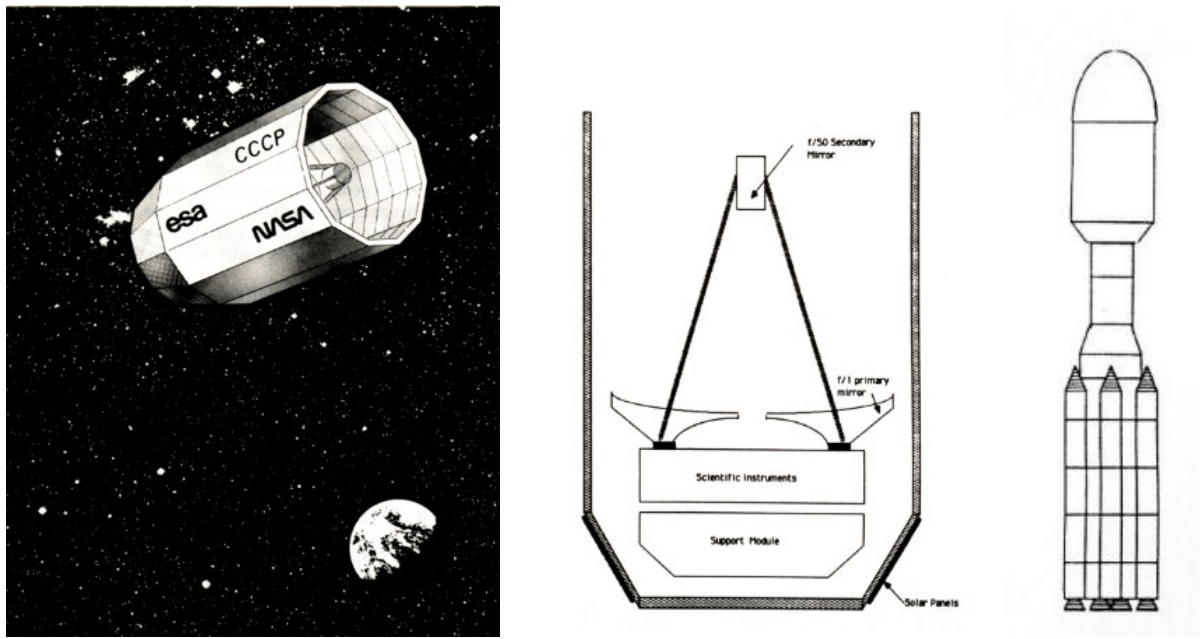

**Fig. 8**. The 10-meter space telescope concept developed by Bely and Roddier in 1986. At left, artist's view of the telescope in geosynchronous orbit. At center, schematic section of the observatory. At right, the USSR's heavy launcher Energia with an enlarged fairing to accommodate the observatory. (STScI)

Although this concept was still very sketchy, it helped persuade those already contemplating future space telescopes that a super HST was in the realm of possibility. STScI had a scientific staff with a broad range of expertise, and informal discussions occurred internally with numerous ideas surfacing about science opportunities for such a large, cold telescope. These were summarized in an invited talk by Illingworth at the General Assembly meeting of the International Astronomical Union held in August 1988 in Baltimore[27] which piqued the attention of the larger astronomical community.

As interest in a successor to HST grew, the project needed a name. Some years previous, advocates for the new type of large telescopes on the ground had called them the New Generation Telescopes (NGT)[28]. Bely proposed the obvious "NGST", for Next Generation Space Telescope. That stuck, and it became the official name of the project during the study phase for the next 15 years. Then, in 2002, it was officially re-christened JWST in honor of James Webb, the NASA Administrator at the time of the Apollo missions.

## 4. A crucial workshop

In 1984, NASA had asked the Science Board of the US National Research Council to make recommendations on the initiatives that NASA should take in space astronomy for the two decades spanning 1995 to 2015[29]. Among the recommendations of the Task Group on Astronomy and Astrophysics published in June 1988 was an 8-to-16-meter telescope destined to further the observations made by HST.

Giacconi had been a member of this task group and, as director of STScI, was positioned to help initiate efforts on that HST follow-up mission. Bely and Roddier had taken a key step in defining NGST, but there was a need to better identify the scientific potential, consult with industry about technical approaches and feasibility, obtain the support of NASA and gain visibility with the relevant astronomy community. All that could be accomplished by convening a joint NASA/STScI workshop, and Giacconi asked Illingworth to organize it at the Space Telescope Science Institute in Baltimore. A three-day meeting ensued in September 1989, attended by 130 astronomers and engineers from government, industry, and universities[30].

There, the goal of a 10-meter free-flyer or 16-meter lunar telescope with capability from the UV to the mid-IR was recognized as being within the then current state of technical development.

It was also agreed that a true successor to HST should not just cover the mid-IR but cover it at high sensitivity. This was vital to the study of the early Universe because the expansion of the Universe shifts radiation from primeval objects out of the ultraviolet and visible bands into the infrared. And that meant the observatory would have to be much colder than initially envisaged (Sect. 3) so as to reduce self-emissivity of the optics to below that of the celestial background. In space, this is dominated by "zodiacal light", the scattering of sunlight by dust in the Solar System[31] (Fig. 9).

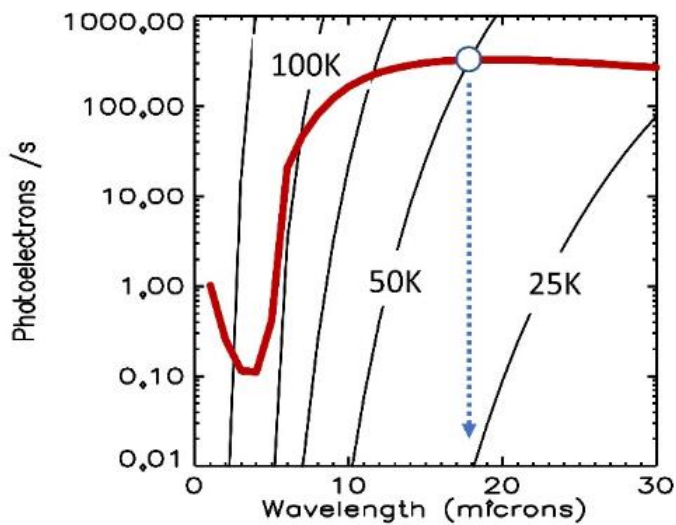

**Fig. 9.** The temperature of the observatory optics should be as low as 50 K for their self-emissivity (black curves) to be less than the zodiacal background (red curve) well into the mid-IR (based on imaging with a 20% bandpass). It is important to note that the optics may not be the only source of instrumental background - see Sect. 10. (Bely/STScI)

Another conclusion reached at the September meeting was that a serious study of the new observatory should begin as soon as possible to avoid a large gap between the end of HST's lifetime and the launching of its successor.

## 5. Detour via the Moon

On July 20, 1989, the 20th anniversary of the Apollo 11 Moon landing, President George H. W. Bush announced his new Space Exploration Initiative with a return to the Moon[32].   Could observing from the Moon benefit astronomy?  Could the successor to HST be a large telescope on the Moon instead of a free flyer?  The subject had been discussed at the 1989 Baltimore meeting and a full workshop was dedicated to these questions in February 1990[33].  So, for a while, one version of NGST was a 16-meter telescope on the Moon. In order to help those in charge evaluate the possibility of a lunar observatory, Max Nein and colleagues from MSFC[34] and Bely[35] briefly investigated what it might look like (Fig. 10 and Fig. 11).

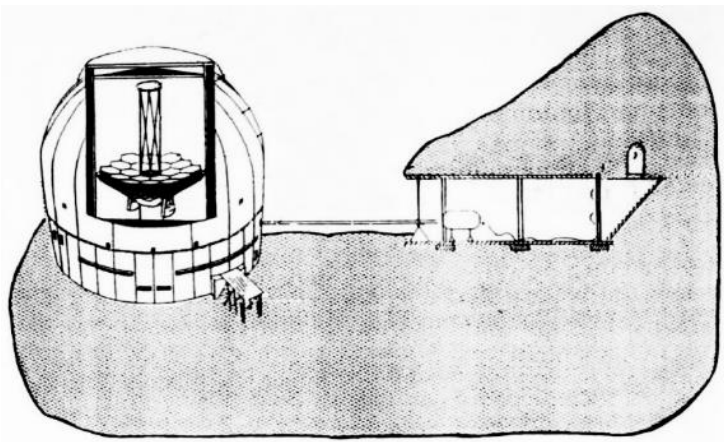

**Fig. 10.**  Sketch of a 16-meter telescope on the Moon, proposed by MSFC.   Instruments and staff are in an underground facility to protect them from cosmic rays. (MSFC)

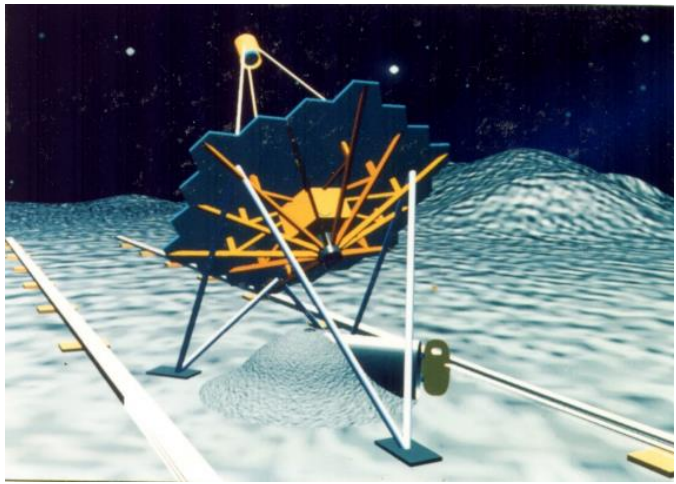

**Fig. 11**. Artist's view of a 16-meter telescope on the Moon, proposed by Bely.   Telescope pointing made use of a hexapod with linear actuators.  During the lunar day, a shield was to be rolled over the telescope to protect it from heat coming from the Sun and the surface of the Moon.  (Bely/D. Berry/STScI)

Sketchy as they were, these studies quickly showed that observing from the Moon is not without problems.  The Moon might appear to combine the advantages of ground-based astronomy with those of space. Like Earth, it offers a stable platform and, like space, it is free from the effects of seeing and opacity as it has essentially no atmosphere.  In addition, if one assumes a manned lunar base, repairing and upgrading the telescope and instruments would be possible.

But the Moon has some serious disadvantages for astronomical observations.  Unlike a free-flying observatory in high orbit which can be protected from Sun and Earth radiation, an observatory on the Moon would be swamped by stray light and be prohibitively hot during the lunar day (when the Moon's surface temperature reaches 400 K). Shielding could not adequately protect the telescope and, in practice, observations during the lunar day would be impossible. Additionally, the enormous temperature differences between lunar day and night would severely tax optical alignments and mechanisms.

Nor is the Moon ideal for infrared astronomy. Although nighttime temperatures are relatively low (soil at 100 K), they still fall short of the very low temperatures (~50 K) passively available in space. One approach for alleviating these temperature issues would be to site the telescope in a crater at a lunar pole. But that would limit sky coverage and would require having a base located nearby.

Furthermore, although gravity on the Moon is only one-sixth that on Earth, its effects are still significant for very large telescopes when changing attitude during observations. Other disadvantages of the Moon include dust, vibrations/moon-quakes, and the fact that less mass can be put on the Moon's surface with a single launch than can be put into space.

Finally, automatic deployment of an observatory on the Moon's surface would be extremely difficult. Prior site preparation and human-assisted installation would appear to be essential, so any major lunar observatory would have to wait for establishment of a manned base, too far in the future for a successor to HST.

In any case, the lunar option for NGST died in 1992, when NASA abandoned plans for any return to the Moon in the near term.

## 6. A minimalist NGST

A flurry of task groups and meetings took place in the 1988-1996 period to formulate scientific objectives and concepts for future space observatories[29,36,37]. In one of the late 1990 Decadal Survey panel reports[38] a new idea was floated: construct a 6-meter space telescope as a first successor to HST, then follow it up with a final, 16-meter telescope on the Moon.

From an engineering point of view, six meters was a welcome size for HST's successor. Both the cost and the difficulty of fabrication of a telescope increase quickly with the diameter[39], so it is always a challenge to design an instrument whose diameter exceeds that of its predecessor by over a factor of 2. Historically, ground telescopes had been growing by such measured steps, from the 60-inch to the 100-inch on Mount Wilson in the early 20th century, to the 200-inch (5-meter) on Mount Palomar in 1950, and to the 8- and 10-meter instruments of the 1980's and 1990's. A 6-meter space observatory would not require the exceptional technological advances needed for the big jump from HST (2.4 meters) to 10 meters.

In addition, being able to launch any space system that is fully assembled and tested is an enormous advantage: there is no risk of deployment mishaps and no need for major adjustments once in orbit. Unfortunately, payload size and mass do matter. Space Shuttle payloads were limited to about 4 meters in diameter. And, although a new family of launch vehicles was under consideration, the Advanced Launch System (ALS) and the Heavy Lift Vehicle (HLV), there was no assurance that their most capable variants would be developed in time to launch an immediate successor to HST as large as 8-10 meters. Hence, 6 meters was seen as having a greater chance of matching available launch capabilities.

Two 6-meter telescope concepts were presented at the 1991 Astrotech 21 workshop by MSFC[40] (Fig. 12) and Bely[41] (Fig. 13). Both had wavelength coverage from UV to 10 μm thanks to operating temperatures of less than 100 K obtained by radiation to deep space, and both were located in a High- Earth-Orbit (HEO – e.g., 100,000 km altitude) to avoid Earth's heat and straylight and to improve observing efficiency. And both assumed that an advanced launching system with a payload fairing on the order of 7 meters in diameter would be developed in the near future.

The MSFC version used a segmented fused silica f/1.4 primary mirror with active control. The observatory's sun side had insulation and a sunshade, the other side was open to deep space for passive cooling. Viewing was limited to a 60-degree band perpendicular to the sun line. A preliminary thermal analysis showed that the optics would reach about 70 K 30 days after launch.

Bely's version emphasized simplicity in order to permit delivery of HST's successor by 2005, that being the end of HST's lifetime as envisioned at the time (14 years).

The primary mirror was monolithic, made of zero expansion glass lightweighted to 80 %, with a conservative f/2 focal ratio to minimize polishing difficulties. Figure errors due to gravity release and thermal effects were corrected by actuators on the back of the mirror. Bending moment actuators[42] rather than piston actuators were used to avoid the need for a backplane support structure.

In order to maximize cooling by increased exposure to deep space, the telescope had no outer baffle. The insulated flexible sunshield which was to open after launch was relatively small, thus minimizing deployment complexity. Although this small sunshield limited viewing to 30 degrees of the anti-sun, it still allowed a sky coverage of 50% over the entire year.

Neither of the two concepts was retained for further study, as the funding from Astrotech21 dried up. But these exercises were useful for investigating approaches to ultra-light active primary mirrors and passive cooling via open telescope structure.

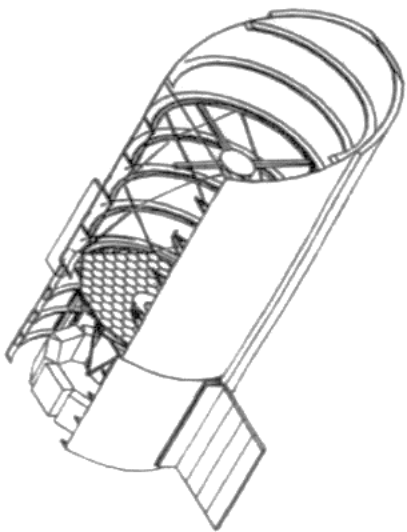

**Fig. 13**. MSFC's concept for a 6-meter telescope in High Earth Orbit. (MSFC)

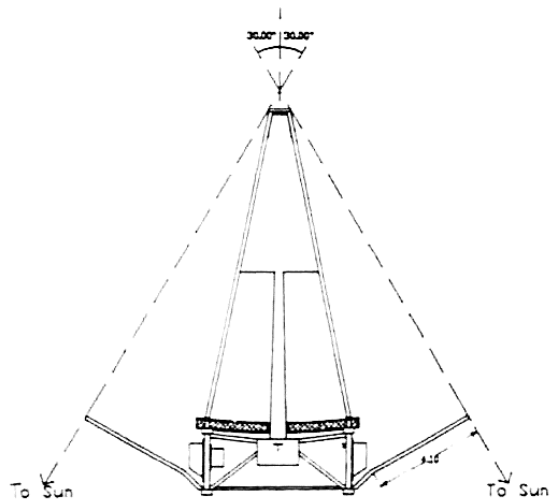

**Fig. 13**. A 6-meter telescope concept by Bely.

## 7. Interlude : The 4-meter "High-Z" telescope

High-Z was a mission proposed to NASA as an Advanced Mission Concept in 1994 by STScI, with Stockman as P.I. as part of NASA's response to the 1990 Decadal Survey recommendation for a space IR mission. Although the main goal was the observation of highly redshifted distant galaxies (with a z of 10 to 15), it was also a powerful general-purpose near- and mid-infrared observatory and thus in the realm of NGST.

The telescope was more modest with a primary mirror only 4 meters in diameter, but its sensitivity was enhanced by being placed in an elliptical heliocentric orbit, taking it to 3 Astronomical Units (AU) from the Sun where the IR zodiacal light background is much reduced (Fig. 14).

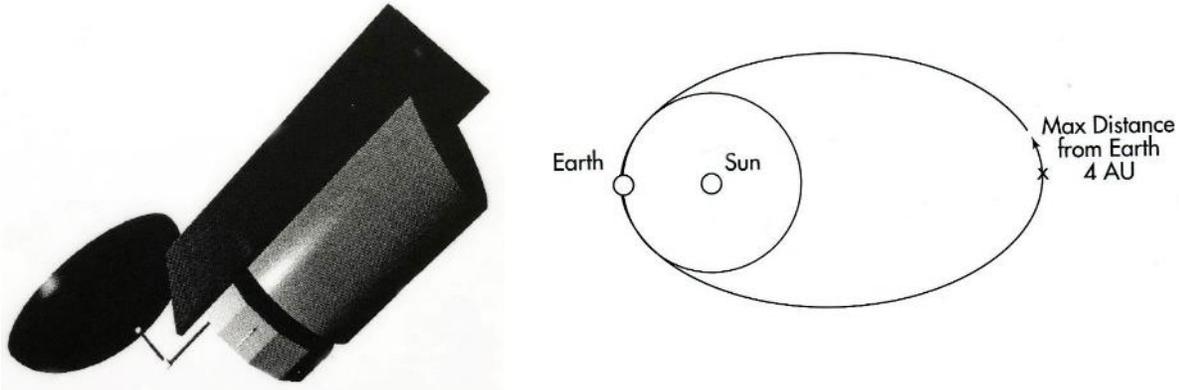

**Fig. 14.** Artist's view of the High-Z mission and, at right, its 1x3 AU orbit, which has a period of about 2.8 years. (Stockman STScI)

The telescope was fully baffled but passively cooled to about 70 K when at or close to 3 AU thanks to reduced solar input, a double sunshield and a single isolating shell surrounding the optics, as per the concept proposed by Hawarden[43]. This was a key difference with SIRTF (later renamed the Spitzer space telescope) which was small enough (85 cm primary mirror diameter) to be cooled cryogenically. Instantaneous sky coverage was limited to a 20-degree band perpendicular to the sun line, but the entire sky could be covered during half of the orbital period.

The mirror was a thin, fused silica meniscus supported by 400 actuators mounted on a stiff backplane support structure, along the lines developed by the Adaptive Large Optics Technologies (ALOT) declassified military program[44]. The observatory could have been launched, fully assembled and tested, by a Titan or Ariane 5, but the proposal was not selected. The competing SIRTF proposal from JPL won.

## 8. NGST on the starting block

On April 24, 1990, just a few months after the 1989 STScI workshop, HST was sent flying into space. Finding a way to correct its faulty optics immediately became a priority for NASA and NGST was relegated to the back burner, particularly when the AstroTech21 program was suspended in 1992. But once HST's optics were corrected in late 1993 the results were so extraordinary that they immediately enthused the general public as well as astronomers. Soon a successor was being discussed by a scientific committee led by Alan Dressler[37]. Its preliminary recommendations included the development of a space observatory of aperture 4 meters or more, optimized for the near-IR. And Edward Weiler, Chief Scientist for HST at NASA, being cognizant of this, seized the occasion to relaunch the study of a successor. In October 1995 he assembled a small team of astronomers and engineers from GSFC, MSFC and STScI to define its scientific goals, formulate a conceptual design, and establish feasibility.

STScI was to concentrate on the scientific objectives, performance requirements, and science instruments; MSFC would focus on mirror and spacecraft issues; and the GSFC team would be responsible for the overall conceptual design. At first, the GSFC team was quite small, with John Mather as Project Scientist, Bernie Seery as Project Manager and a few scientists and engineers. These included Richard Burg, Charles Perrygo, Eric Smith, and Pierre Bely, who had been reassigned there as Mission Architect, and later Paul Geithner. The STScI team, headed by Peter Stockman, included Chris Burrows, Mark Clampin, John MacKenty, Larry Petro and Massimo Stiavelli. The MSFC group included Jim Bilbro and Max Nein. These teams grew over the next four years, adding part-time personnel and drawing on many scientists and engineers from NASA, particularly Richard Capps, Dan Coulter and Dave Redding from the Jet Propulsion Laboratory (JPL), and a few engineers from the industry acting as consultants, in particular Mike Krim of Perkin Elmer and Robert Woodruff of Ball Aerospace (Fig. 15).

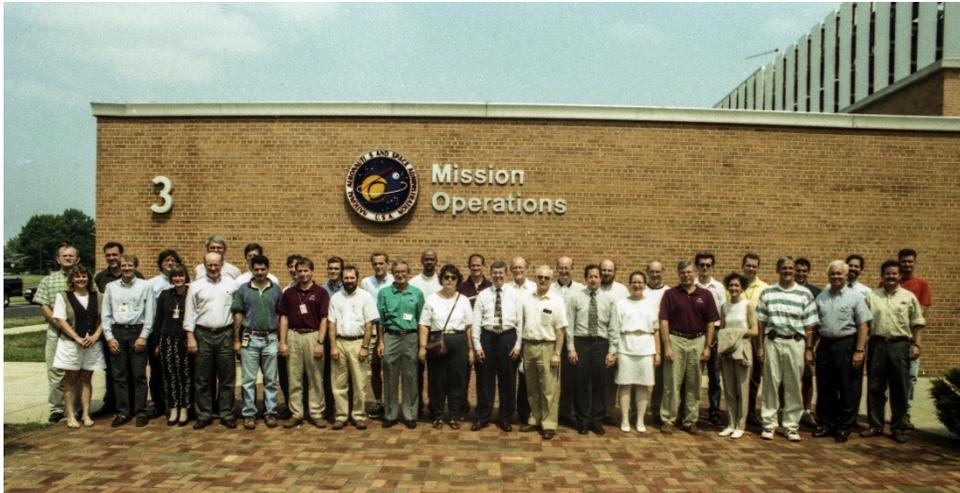

**Fig. 15.** The NGST team at the Goddard Space Flight Center near the end of the NGST study in August 1998. Some members from MSFC and STScI were there, but not all. Project Scientist John Mather and JPL members were absent. (Photo Credit: NASA)

## 9. The Eureka moment.

In January 1996 the embryonic NGST study team began discussing architectural options. The optics, science instruments and space support systems could follow then-current practices or recent new developments. The unresolved issues were how to make the observatory extremely cold and what would be the best orbit (Fig. 16).

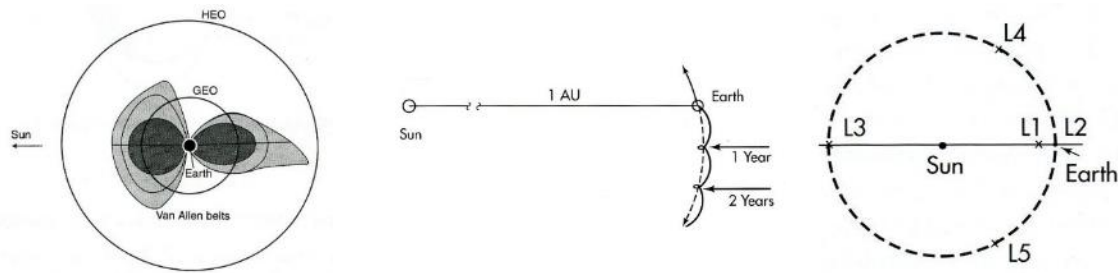

**Fig. 16.** The locations favored for NGST included a circular high Earth orbit (around 100,000 km altitude to be beyond the Van Allen belts) where the Earth subtends an angle of only 7°, but which is energetically costly (left); a drift orbit (center) which is energetically economical but suffers from steadily increasing distance to the Earth, limiting mission time; and one of the Lagrange points of the Sun-Earth system (right) where the centrifugal force acting on the object exactly compensates the attraction of the Sun and Earth, especially L4 and L5 which are fully stable, and L2 which is metastable and requires station-keeping.

The Eureka moment occurred in late February 1996, following a brainstorming meeting at GSFC where all options and ideas had been on the table: telescope configuration and deployment, several types of orbits, approaches to passive cooling, science instrument complement, etc., but no clear solution to the puzzle had emerged. Stockman and Bely were driving back to STScI on the Baltimore-Washington Parkway. It was rush hour and Stockman was at the wheel, responding to Bely who was recapitulating the various arguments from the meeting's discussions and trying to extract a rationale. Things went something like this:

1. For such a large telescope, passive cooling is a must, and we need to be far from Earth to avoid being swamped by its heat.

2.  If we are far from Earth, why not eliminate the external tube and have the telescope completely naked? Radiative cooling is very ineffective at extremely low temperatures and the larger the view of deep space, the better.

3.  Without an external tube, light from off-axis stars could be eliminated by internal baffles, and a sunshield would take care of the Sun. But we would still need protection from the light of the Earth and the Moon.

4.  In a High-Earth Orbit, an Earth-trailing orbit, or at the stable Lagrange point No 5, the sunshield could also be used to block the Earth and the Moon, but this would entail serious viewing constraints.

5.  We keep rejecting L2 because it is not a fully stable location. But there, the Earth and the Moon are all in the same direction… and would be hidden by the sunshield… It is really an ideal site for NGST - actually, the very best one in the vicinity of Earth[*].

6.  But that would mean putting a propulsion system on the observatory for station-keeping, and also a short mission time, limited by propellant.

7.  Well, that is the price to pay, but so obviously the best solution…

Now, these principles remained to be materialized in an observatory concept, and that happened within a few days. Although, a 4-meter primary mirror had been the starting baseline of the NGST study, the goal for HST's successor had always been larger. Mather and Seery were hoping for 6 to 8 meters and NASA Administrator Dan Goldin urged for that range during his speech at the American Astronomical Society (AAS) in January 1996.

Ariane 5, one the largest launchers at the time, had a payload envelope diameter of 4.5 meters. How big a telescope could possibly fit inside? After considering possible configurations, Bely selected an "open flower" concept modeled on the High Angular Resolution Deployable Interferometer (HARDI) that he had developed in 1989 with Burrows, Roddier and Gerd Weigelt for a potential ESA mission[45]. The primary mirror had a 3-meter hexagonal center core surrounded by 6 panels folded upward for launch, an overall diameter of 7 meters upon deployment, and an area equivalent to that of a 6-meter filled circular aperture (Fig. 17).

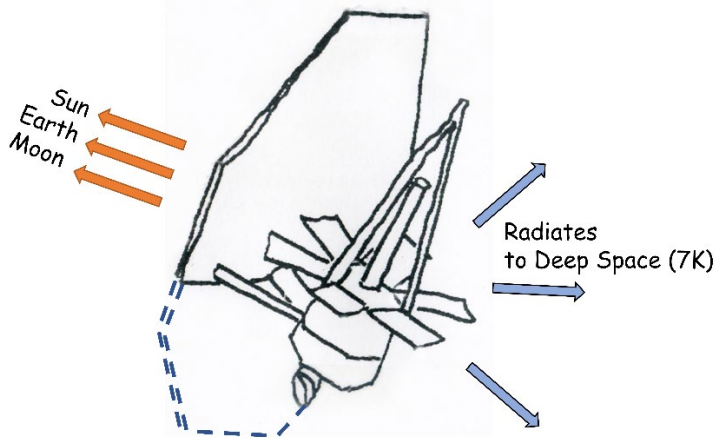

**Fig. 17.** Sketch of the concept proposed for NGST in February 1996.

A crude model of the concept was circulated and generated much interest. Mather remarked that the sunshield would have to be extended at the bottom to balance solar pressure and allow for greater pitch.

And that was it. Many changes would still be made, but the basic architecture of NGST was born.

---

[*] Arguably, the best infrared observing site in the entire solar system may be the 2nd Lagrange point of the Sun-Jupiter system. Because of Jupiter's huge mass, the L2 point is close to the planet and always in its shadow. There, in total darkness, with non-solar nuclear power, an observatory could reach temperatures as low as 7K, benefit from lower zodiacal background and have a sky coverage close to 100% (from a presentation to the NASA Administrator by Petro and Bely in 1999).

A major observatory with a deployable and naked telescope, a propulsion system, a short lifetime because of propellant limitations and located so far from Earth sounded far-fetched at first. But the advantages of the concept were so compelling that it immediately became the basis for the NGST study.

At that time no spacecraft had been located at L2. But two months later, in April 1996, the NGST team was comforted in that choice when they learned that the Microwave Anisotropy Probe (MAP, now WMAP) had been selected by NASA as a MIDEX mission and would be sited there. MAP instruments had to operate at about 90 K and, like NGST, would benefit from being at L2 for radiative cooling with combined protection from the Earth and the Sun. But for MAP, the primary reason was to minimize magnetic, thermal, and radiation variations from the Earth and Sun that affect the science data[46,47].

## 10. The NGST "Yardstick" concept

With the site and basic architecture defined, fleshing out the NGST concept started in earnest, this time with a goal of 8 meters. A first-order design was essentially in hand by May 1996 and was presented at SPIE conferences in June and August 1996[48,49]. The concept shown in Fig. 18 was referred to as "Strawman" or "Yardstick", since its purpose was to establish first order feasibility and serve as a reference design that industry proposals could later be compared with.

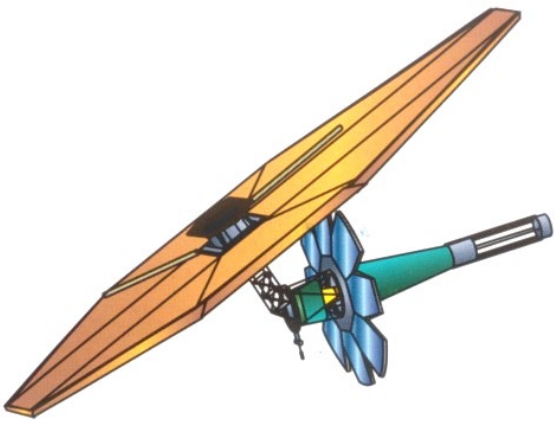

**Fig. 18**. Artist's view of the GSFC "Yardstick" concept. The sunshield was composed of 6 parallel membranes to be deployed by inflatable booms. The spacecraft support module was located on the warm side of the sunshield and kept at room temperature for the proper functioning of mechanical and electronic equipment. (NASA/GFSC)

While HST and most modern telescopes use a Ritchey-Chretien optical configuration to provide excellent images over a wide field with only 2 mirrors, a Korsch 3-mirror system[50] was selected for NGST. A third powered surface allowed for a fast primary mirror to reduce telescope length. It also provided a real, accessible pupil where a deformable mirror (DM) could be located (Fig. 19). The deformable mirror, equipped with 349 actuators, corrected primary mirror figure errors due to gravity release and thermal effects, allowing the telescope to be diffraction limited at 2 μm. Finally, a fast-steering mirror (FSM) was in the train for fine pointing

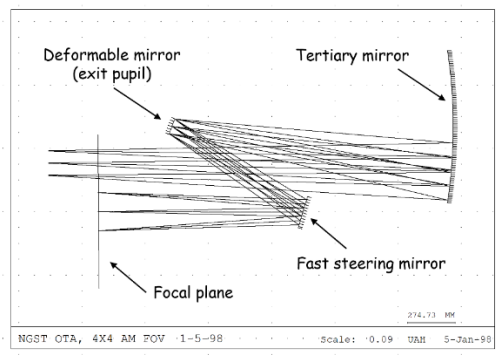

**Fig. 19.** Close-up view of the tertiary-to-focal plane optics.

The primary mirror was composed of a central core and 8 petals folded alternatively up or down for launch, similar to a proposal by Perkin-Elmer's Mike Krim at the 1989 NGST workshop at STScI[30] (Fig. 20). With an overall diameter of 8 meters, the total collecting area was equivalent to that of a full circular aperture 7.2 meters in diameter. The secondary mirror sat atop a sliding pod inside the inner baffle and was deployed after launch.

Each optical element was mounted on 3 actuated bipods providing 6 degrees of freedom for on orbit adjustment. All mirrors were cryo-null-figured, i.e. figured at room temperature, tested at cryogenic temperature, then refigured at room temperature to correct errors detected during the test. The primary mirror segments had no shaping actuators, relying on the deformable mirror to correct their on-orbit deformation. After launch and cooldown, a cascade of image-based methods was used to capture, align and phase the primary segments. Once phased, the thermally stable architecture of the Yardstick, augmented by occasional image-based wavefront sensing and control, ensured a high degree of wavefront stability.

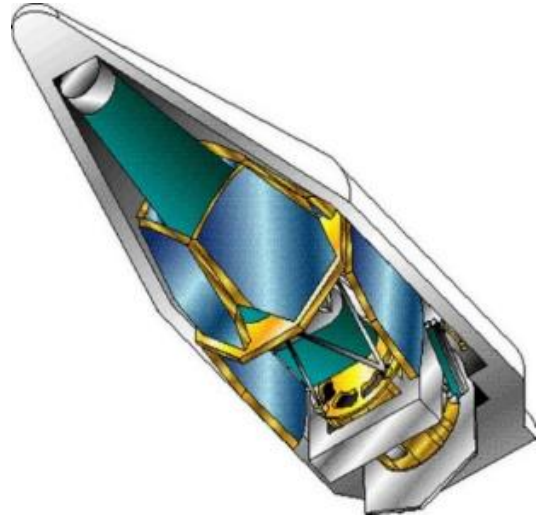

Spectral coverage extended from 0.6 to 28 μm. Coverage of the visible was not considered essential since that was already available with HST and atmospheric seeing correction with very large ground telescopes was steadily improving. The upper limit of coverage was determined by the capability of the Si:As detectors.

**Fig. 20.** The telescope, space support systems and sunshield packed for launch in an Atlas II ARS fairing. (NASA/GFSC)

All optics were coated with gold, which provides excellent reflectivity in the infrared and still has good reflectivity down to 0.6 μm (Fig. 21). Opting to drop spectral coverage in the lower part of the visible for NGST coupled with gold coatings increases the sensitivity of the telescope in the near-IR, which is equivalent to having a significantly larger telescope.

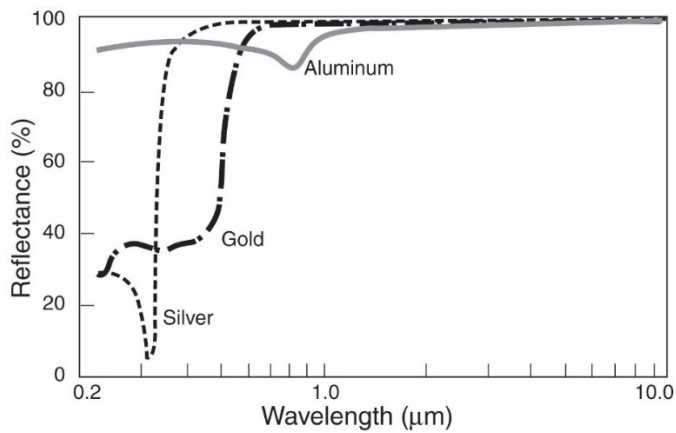

**Fig. 21.** Reflectance of common mirror coatings. Silver is seldom used because it requires protection against oxidation, leading to losses at some wavelengths due to destructive interference in the overcoat. Aluminum is the best general-purpose coating in the visible and IR, but gold is significantly better in the near-IR up to 2 μm, and about 1% better beyond that.

Beryllium was used throughout the optics and structure to minimize mass and misalignments during cooldown. Beryllium is the ideal material for cryogenic space telescopes because of its high stiffness-to-density ratio and high thermal conductivity. It also has the advantage of possessing a very low and uniform coefficient of thermal expansion at cryogenic temperatures.

The sunshield consisted of six reflective layers deployed by four axially extendable mechanical or inflatable booms[51]. To avoid attitude torque, the center of solar pressure was aligned with the system's center of mass for nominal orientation normal to the Sun. The diamond-shaped sunshield was sized to allow the observatory to be pitched +/- 25 degrees with respect to the sunline, providing an instantaneous sky coverage of 40% of the entire sky, and full sky coverage over about six months.

One problem with a naked telescope is that it is not protected from surrounding sources of emission, as it would be if encased in a tube. Direct illumination of the focal plane by off-axis sources is prevented by baffles, one around the secondary mirror, one above the primary mirror, and by a field stop at the first focus.

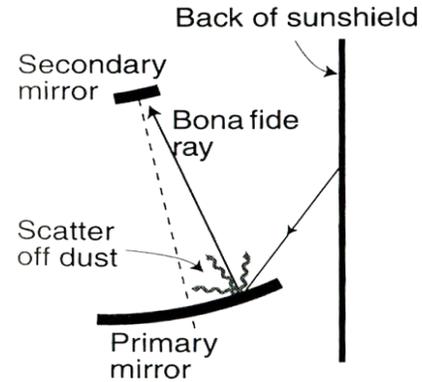

However, the non-optical surfaces of the observatory and, in particular, the large sunshield's back surface can still irradiate the focal plane by scattering off the primary mirror. By radiating to deep space, the optics of a fully open telescope can be cold enough to reduce self-emission, helping to minimize the background in the infrared. But radiation from the back of the sunshield scattering off the primary mirror (Fig. 22) can become the dominant instrumental background if the back of the sunshield is not cold enough[52].

**Fig. 22**. Scatter from the sunshield is the main source of instrumental background in a "naked" space infrared observatory.

Another issue with the whole design was that, with a large flexible sunshield and an ultra-light telescope, the observatory was very flimsy. Consequently, maintaining the optical line of sight on a scientific target by "body pointing", i.e., direct pointing of the whole telescope as was done with HST, would be almost impossible. The solution was to use a fast-steering mirror (FSM) in the optical train to stabilize the line of sight.

To estimate the performance of an observatory it is standard practice to use the "root mean square" error budgeting method[53], as had been done for HST. This was insufficient for the NGST Yardstick because of the complexity and interdependence of its systems. Instead, an integrated end-to-end computer model of the entire observatory, the NGST Mission Simulator, was developed building on commercial and JPL analytic tools. It included the optical, thermal, structural and attitude control systems and was used to demonstrate that the architecture met its scientific requirements and was resilient to post-launch misalignments and uncertainties[54].

Although by 1997 industry had begun working on the NGST project, the Yardstick concept continued to be refined for several more years, taking advantage of all the talent and experience possessed by the assembled team (the complete list of members is given in Appendix A of Reference 70). Detailed analyses were performed on specific issues and reported in several papers [55,56,57,58]. A series of internal monographs was also published at GFSC to capture the rationale behind the options adopted for the Yardstick architecture and to summarize the feasibility study that had been done (Table **1**).

**Table 1**. The NGST monograph series published by the NGST Project Office at GSFC (1998-2001).

| Monograph | Title | JWST Report # | DOI # |
|---|---|---|---|
| 1 | NGST Yardstick Mission | 474 | |
| 2 | Straylight Analysis of the Yardstick Mission | 367 | 10.13140/RG.2.2.13188.13443 |
| 3 | Implication of the Mid-infrared capability for NGST | 478 | 10.13140/RG.2.2.19374.73282 |
| 4 | NGST Integration and testing strawman plan | 588 | |
| 5 | System level requirement, recommendations and guidelines | 587 | 10.13140/RG.2.2.21052.45449 |
| 6 | NGST performance analysis using integrated modeling | 1794 | |
| 7 | NGST Optical quality guidelines | 627 | 10.13140/RG.2.2.27763.34086 |
| 8 | The radiation environment for NGST | 673 | |
| 9 | NGST optical and system testing strawman plan | | |

Eight additional monographs titled *Sunshield design*, *Estimate of meteoroid impacts*, *The integrated science instrument module*, *Optimal primary mirror actuator placement*, *Wavefront error budget*, *Field required for guiding, Optimal selection of observatory parameters*, and *The L2 environment* were initiated or planned but, unfortunately, never completed. The Monographs are mentioned here because, completed or not, they illustrate the specific subjects which, at that time, the designers felt it necessary to study thoroughly to ensure the soundness of the proposed architecture.

Monographs fill a particular need. Bulletized presentations are typically high-level simplified descriptions with short lives, and published papers give a good summary of the major subjects. But technical reports, or monographs, are precious because they are more detailed and exploratory, and address more subtle aspects of a project. In the heat of the conceptual phase, they force designers to clarify rationales, issues and proposed solutions. And for contractors in charge of final designs as well as for designers of similar future missions, they supply an in-depth record of what has been studied and can serve as a basis for their own work.

It is worth noting that government institutions and universities provide a work environment with enough freedom to do detailed exploratory studies such as the Yardstick, which can then serve as a firm basis for industry, with its ability to bring in diverse resources and experience, to start working on the next steps.

## 11. Optimizing the NGST architecture for scientific performance

The Yardstick study had produced a feasible architecture, but it needed to be optimized for scientific performance. Scientific goals can be translated into basic requirements such as sensitivity, angular resolution, wavelength coverage, spectral resolution, field of view, etc. These are often in conflict, however. For example, there are different requirements for the study of high redshift supernovae and for high redshift galaxies. For the supernovae studies, the rarity of the events requires a large field of view, while moderate angular resolution is often sufficient since these are point sources. High redshift galaxies, on the other hand, are numerous and small but extended, and therefore require exquisite angular resolution for detailed analysis of their morphology. Another example is the conflicting demands on general purpose observatories that are required to observe nearby, relatively bright objects as well as obtain deep exposures of the distant Universe. A practical compromise must be found, and this can be done by "weighting" the various scientific goals and their corresponding observational requirements. An essential tool for accomplishing this optimization is the ``Design Reference Mission'' (DRM). This concept has its origin in space programs, where it serves as a basis for simulations to validate the hardware of a mission. The idea behind the DRM is to define a strawman observation program and run it against a mathematical model of the observatory. For NGST, the strawman program was defined by STScI[59], relying mostly on the scientific recommendations of the HST & Beyond report[37,70]. It was then run in the observatory simulator to optimize telescope and science instrument temperatures, field of view and spectral coverage and to study how much descoping could be tolerated, if necessary[60] (Fig. 23).

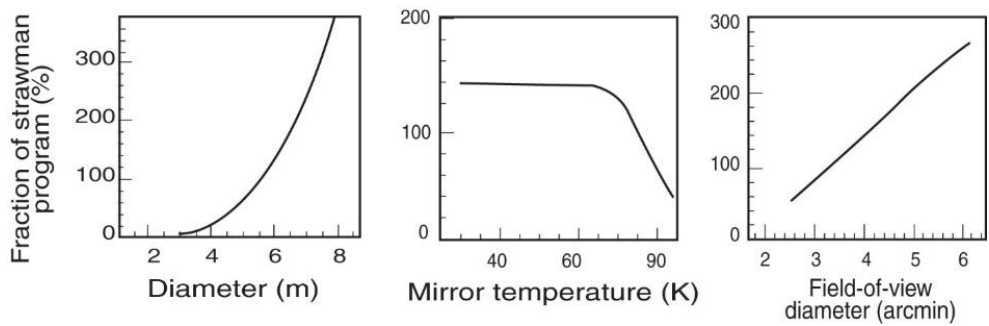

**Fig. 23.** Completion rate of the NGST core scientific program for a mission duration set for 5 years, as a function of telescope diameter (left), mirror temperature (center), and field of view (right). These plots illustrate the rapid increase in program completion with increased telescope diameter and the dramatic decrease when the temperature of the optics exceeds 70K. (Stiavelli/STScI)

This parametric analysis showed that, in order to satisfy the mission goals defined in the DRM, the primary mirror should be at least 6 meters in diameter, the mirror temperature under 70 K and the field of view about 4 arc minutes.   A study using the project Science Working Group's DRM and the mission simulator reached similar conclusions[61].

Defining a new, more capable observatory on the basis of the desirable scientific goals of the moment has inherent limits, as explained in Sect. 2.   But the DRM exercise was still useful by helping to reach a reasonable balance among the main observational parameters.   It also illustrated how sensitive the scientific potential of the observatory was to the mirror's diameter and temperature, and this independently of the research theme.

## 12. Last attempt at a monolithic primary

Although the NGST baseline with a segmented and deployable primary mirror was attractive, and large segmented mirror technology had recently been successfully demonstrated on the ground by the 10-meter Keck telescope[62], the system was still considered relatively complex and delicate for space applications. On the other hand, fabrication and polishing of large monolithic mirrors of the 8-meter class was established technology, as shown by several ground telescopes under construction at the time (VLT, Subaru, Gemini, and Magellan telescopes). So, early in 1996, an alternate design for NGST was proposed, using a monolithic primary mirror.

The launch vehicles available at the time (Atlas and Ariane) were not capable of accommodating a circular 8-meter primary mirror, but an elongated shape could fit.   This is what the alternate NGST concept proposed[63], supported by Giacconi, who had meanwhile become director of the European Southern Observatory (ESO) which was building a set of four 8-meter ground telescopes in Chile.   The proposed primary mirror for NGST was a truncated ellipse 8 meters by 4.5 meters.   It was to be made either of beryllium light-weighted by about 90%, or of a 2 mm-thick figured glass facesheet supported by actuators. The secondary mirror was mounted on an articulated tripod folded for launch and deployed by a simple spring release in one of the legs (Fig. 24).

The advantage of this alternate design was simplicity.   Its main disadvantage was a smaller collecting area, equivalent only to that of a 6-meter circular telescope - a failing grade at that time when 8 meters was the baseline.   And when the later descope to 6.6 meters was made, the final design had advanced too far to revisit this approach.

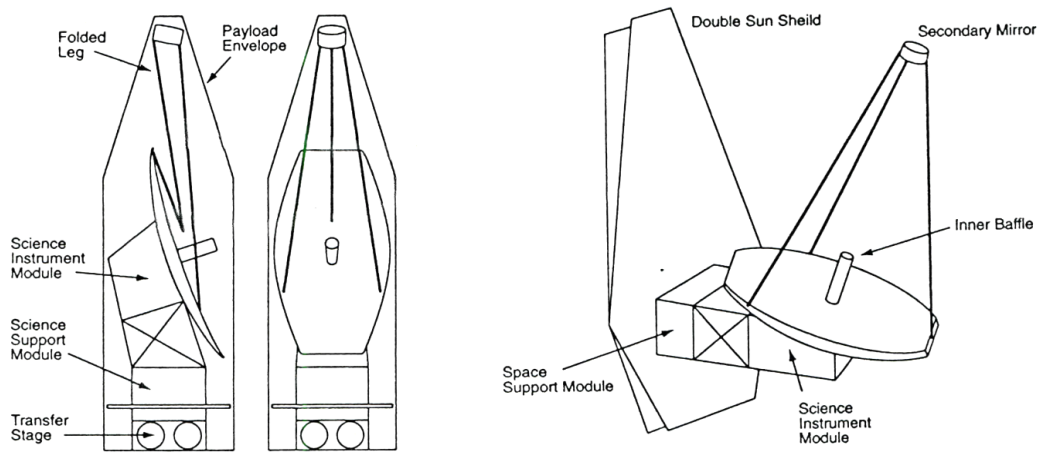

**Fig. 24.** The elliptic mirror concept, shown packed in an Ariane 5 fairing at left and deployed on orbit at right.

### 13.  The "break the cost curve" mantra

Although various diameters had been considered for HST's successor over the preceding 10 years, the baseline for the NGST concept studies was 8 meters.  This was reinforced in Dan Goldin's speech in June 1999 at the 100th Anniversary Meeting of the AAS and soon after was cemented into the National Research Council's 2000 Decadal Survey[64].

Now, whether constructing a house or a large telescope, cost is an integral part of the development of the architecture.  The design must fit within the budget.  In the case of HST's successor, NASA imposed a serious budgetary constraint: it should cost less than HST had.  And what was supposed to make this huge gain possible? New technology.

Technological progress has indeed reduced the cost of products in many fields, significantly so in the case of ground telescopes.  Until the 1980's a well-established power law was that their cost grew roughly as the 2.6th power of the diameter of the primary mirror (Fig. 25).  According to that power law, the 8- and 10-meter class telescopes built in the 1990's would have been completely unaffordable.  But technological improvements had made it possible to significantly lower their cost: the mastering of the aspheric polishing process brought faster primaries, hence shorter telescopes and smaller domes; the use of alt-azimuth instead of traditional equatorial mounts reduced the mass of telescope structures; and the advent of computer systems and active optics made it possible to control the figures of very thin segmented primary mirrors, thus reducing the mass and cost of the optics and supporting structures.

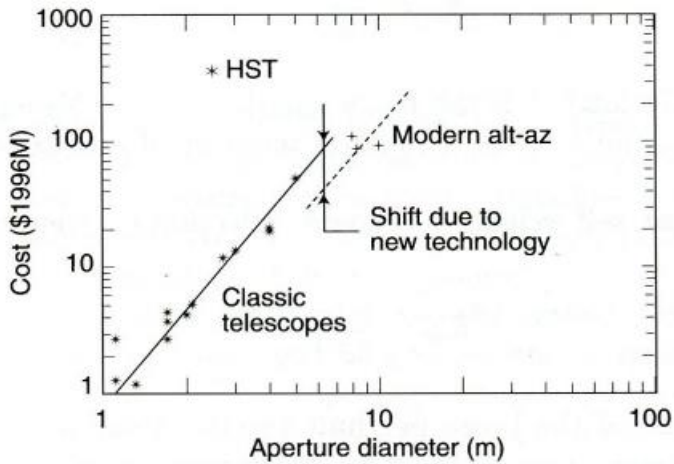

**Fig. 25.** Cost of ground telescopes vs. primary mirror diameter in constant US dollars.  The cost of HST (optical telescope assembly only) is shown for reference. (Bely)

At that time there were not enough data to determine how cost varied with diameter for space astronomical telescopes, but common sense and hearsay suggested a cost-diameter power law on the order of 1.8 or maybe 1.6, the lower exponent relative to the ground 2.6 exponent being justified by the fact that, minus dome, building and gravity, a space telescope's fabrication cost scales, at most, as the aperture area.

Still, even a 1.6 power law meant that an 8-meter space telescope should cost almost 5 times as much as HST.  Hence the mantra: "Break the cost-diameter curve" with new technology. Contrary to popular belief, however, the effect of new technology is not to flatten the cost curve.  The diameter of a telescope's aperture is a fundamental scaling factor which is hard to break.  A bigger something of the same kind will always cost more.  What new technology does is reduce the cost of building something of a given size. It lowers the constant in front of the power law but not the power law exponent. New technology does not break the curve, it shifts it downward.

This is precisely what had happened with the 8-to-10-meter class ground telescopes, and it was hoped that new technology and new architectural and management approaches would do the same for NGST. Gains were expected thanks to segmented optics, active optics, less demanding optical quality by prioritization of infra-red wavelength over UV and visible, and management and programmatic improvements (faster schedule, early technology development)[65].

On the other hand, a number of factors were pulling the other way: cryogenic temperature, deployable optics, deployable sunshield, station keeping, maturing the technical readiness of many break-through technologies (see Sect. 17). Furthermore, NGST was supposed to be ready for launch in 2008, so there was not much time for even fast-paced additional technological developments.

In the late 1990's, the NGST project team was desperately trying to meet the budget limit imposed by NASA[65]. But whether using models or direct first order estimates, it seemed impossible to reach a cost comparable to HST's. One possibility was to restrict spectral coverage to the near-IR. This saved on the cost of the science instruments complement but relatively little on the telescope and sunshield. This was because the near-IR detectors needed to be cooled to about 40K, and it was less risky to stay with the baseline design which provided that passively rather than having to develop a cryocooling system (see Monograph 3 in Sect. 10). But this solution still did not reduce costs enough. So, in June 2000, Bely and Burg made an internal proposal to lower cost and shorten fabrication time by reducing the aperture to 6 meters in addition to eliminating the mid-IR capability[66]. Descope to 6.6 meters would indeed occur later, in the Phase A design phase, though fortunately maintaining the mid-IR capability thanks to the insistence of Seery, Smith and Mather who then declared: "No mid-IR, no mission."

Cost models for space observatories are based on parameters that make physical sense but, as pointed out by Stahl[67], they have serious limitations when applied to observatories of widely different architecture, dimensions, spectral coverage, operating temperature and cost accounting systems. Still, it may be instructive to compare the final cost of JWST to what could be extrapolated from the cost of HST.

Let us attempt it in very rough terms, assuming that the overall cost of the observatory is proportional to that of the telescope proper, that cost-diameter law follows a 1.6 power law and that the cost of the telescope is inversely proportional to $T^{0.2}$, where T is the absolute temperature of the telescope, as suggested by Horak[68]. Extrapolating the 2.4-meter, 290 K HST to the 6.6-meter, 50 K JWST leads to a cost increase factor of about 7.

In 2018 a comparison was made between the cost of HST and that of JWST[69]. JWST was nearing completion then at an expected final cost of $10 billion, while the total cost of HST, adjusted for inflation, was estimated to be $ 9 billion. The final cost of the 6.6-meter JWST was thus fairly close to that of the 2.4-meter HST – nothing like the predicted factor of 7. The cost curve had indeed been shifted downward by a very significant amount.

Focused technological advances were certainly a factor, as NASA had planned, but most of them went to enabling the mission, not to reducing its cost. Other factors were at play.

A significant cost reduction compared to HST stemmed from the elimination of coverage in the visible. This relaxed the optical requirements and enabled the use of more efficient gold coatings. According to Horak's law, the cost of a space telescope is inversely proportional to $\lambda^{0.18}$, where $\lambda$ is the operating wavelength, and setting the diffraction-limited wavelength at 2 μm for JWST leads to a 20% cost reduction for the telescope.

Many other factors played a major role in reducing the cost, including: general advances in all fields over time, computerized design, simple open architecture, active optics, optimization via a DRM, end-to-end modeling, and, importantly, scientists, government and contractors working as a team, brimming with experience, talent, and eager to respond to the challenge.

## 14. The 1996 industry feasibility studies

By May 1996 the GSFC "Yardstick" concept for NGST had sufficiently advanced to be credible and NASA decided to present it to selected aerospace companies in a meeting held at STScI. Following that meeting NASA funded feasibility studies by consortia led by Lockheed Martin and TRW/Ball Aerospace. The results of these studies were presented in late August 1996 and summarized, together with that of the GSFC NGST team, in a seminal report edited by Stockman in June 1997[70].

TRW proposed a concept leveraged from their High Accuracy Reflector Development program (HARD) that allowed an 8-meter class segmented primary mirror to be stowed in a 5-meter class launcher fairing (Fig. 26). The six hexagonal segments of the primary mirror were stacked one on top of the other for launch. For deployment, the stack would rotate as a block around the central segment, drop each segment into position and lock it to its neighbor. The secondary mirror was deployed by unfolding its supporting legs.

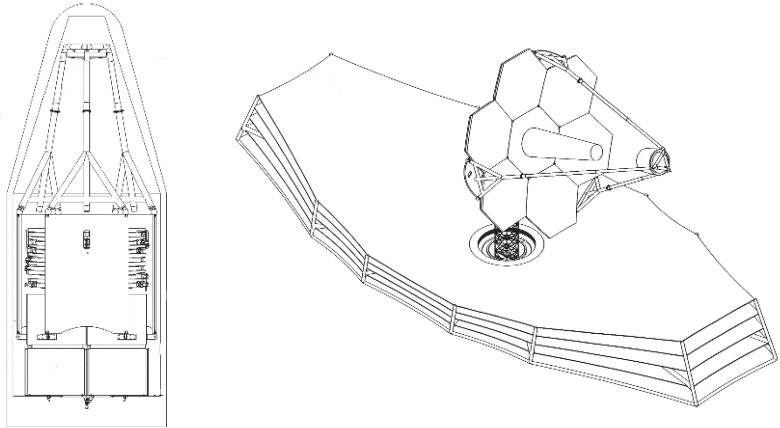

**Fig. 26.** The TRW concept shown stowed at left and deployed at right.

Ball Aerospace proposed stowing the primary mirror "vertically" in the launcher fairing using a two-folding wing geometry as shown in Fig. 27. The primary mirror segments could be either hexagonal or keystone shapes.

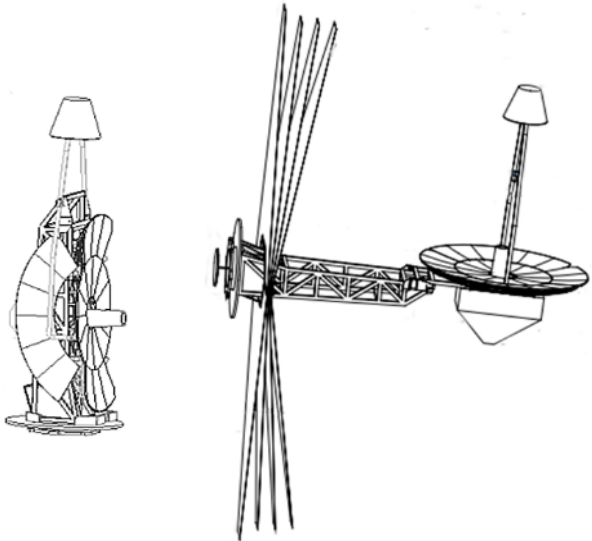

**Fig. 27.** The Ball Aerospace concept, shown stowed at left and deployed at right.

Both the TRW and Ball concepts were sited at the Lagrange 2 point and used multi-layer deployable sunshields to passively cool the telescope and instrument packages, with a tower to provide separation between the telescope and the sunshield. Both concepts used gimbal mechanisms to steer the telescope without having to pitch the whole observatory, thus minimizing the size of the sunshield and keeping it fixed in space, easing the pointing control system.

To avoid the complexity of a deployment, the Lockheed Martin study revisited the High-Z concept (see Sect. 7), using a non-deployed 6-meter monolithic primary mirror with a fixed secondary mirror, as shown in Fig. 28. The sunshield was also fixed, in the form of a half cylinder, leaving the observatory open to deep space for cooling. In order to achieve the IR sensitivity of an 8-meter telescope close to Earth, the observatory would be operated in an elliptical heliocentric orbit with aphelion of at least 3 AU, thus benefiting from lower zodiacal background. Although simple this concept did not offer scalability for larger future telescopes, and it was not certain that launch providers had the market to warrant the development of the required 6-meter fairing.

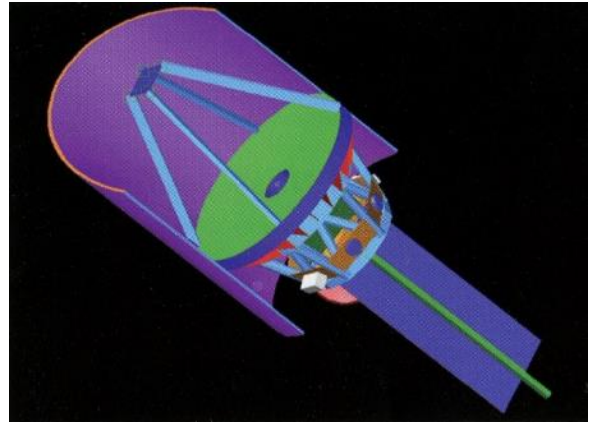

**Fig. 28.** The Lockheed Martin 6-meter monolithic mirror telescope in 1x3 AU orbit.

In order to propose a scalable architecture Lockheed also studied a concept called Multi-Ap. This was an array of afocal 1- to 2-meter telescopes whose beams were combined to yield the equivalent of an 8-meter telescope, as exemplified by Fig. 29. Since the telescope array was compact and not spatially dispersed as in an interferometer, it did not suffer as much energy loss in its PSF, as explained in Fig. 7. The architecture could be scaled for larger systems by adding telescopes, provided that the beam combiner was appropriately designed. However, this concept had a limited field of view and suffered from reduced optical throughput due to the complexity of its beam combiner.

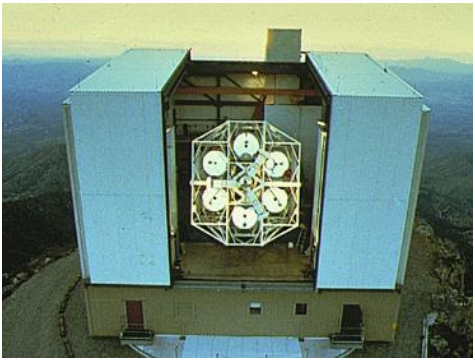

**Fig. 29.** For proprietary reasons no image of the Lockheed Multi-Ap is available. It followed the same principles as the original Multi-Mirror-Telescope (MMT) in Arizona, shown here, which was composed of 6 individual telescopes whose beams are recombined at a single focus. (Steward Observtory)

As we have just seen, inviting the aerospace companies to work on the conception of a successor to HST was extremely fruitful. Applying their extensive and deep experience with the real world of space missions to the "Yardstick" concept, they had been able to produce, in just a few months, a large and diverse set of possible 8-meter class high IR-sensitivity observatories that could be accommodated, both in mass and volume, by then current launch vehicles.

## 15. Nexus – An in-space technology testbed

NGST was to be a novel space observatory – the first large, deployable, passively cooled infrared telescope with active optics ever to be launched. No active correction using on-orbit wavefront measurements had ever been accomplished, nor had a large observatory ever been launched without full scale ground testing. Because of its size it was difficult, if not impossible, to fully test it on the ground and, because of its low operating temperature, difficult to model mathematically.

Minimizing risk would necessitate extensive and costly testing. In mid-1997 Seery, manager of the NGST project, concluded that it would be quicker and more economical in the end to build, launch and operate a scaled prototype. Such a prototype, which Seery dubbed Nexus for being the "link" to NGST, would test the key technologies that could not be fully validated on Earth (Fig. 30).

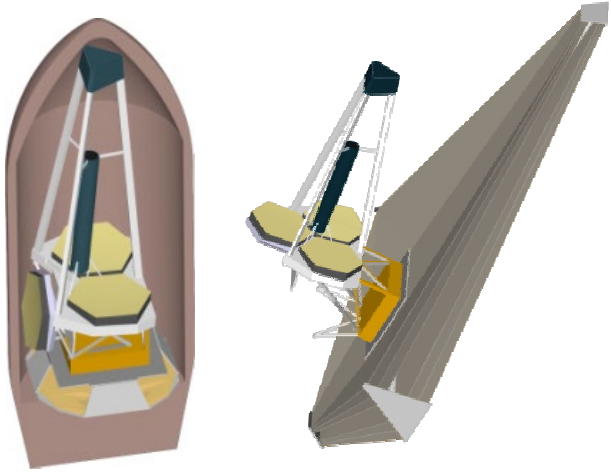

**Fig. 30.** Artist's view of Nexus: stowed for launch at left and deployed on orbit at right. The observatory had a 2.8 m primary mirror consisting of three 1-meter segments, two fixed and one deployable. Like NGST, it was to be sited at the L2 point and cooled to about 50K. (GFSC)

Around that time Ariane 5, which would eventually launch JWST, failed on its maiden flight because of overconfidence, inadequately validated navigation software and insufficient testing. This led Gérard Brachet, one of the key people behind the Ariane project, to reflect humbly: "We should never forget that computer simulations and ground tests cannot answer all questions"[71]. That failure could only increase Seery's eagerness for an in-space NGST testbed. To make the mission even more attractive an appealing scientific observing package was piggybacked onto the purely technical program. NASA eventually agreed to this with a launch scheduled for 2004, in time for the planned NGST Critical Design Review (CDR). However, while having substantial rationale from an engineering perspective, the Nexus program was cancelled in December of 2000. This resulted both from lack of funding and considerable pushback from the scientific community which had come to fear that, if a 3-meter class IR mission such as Nexus were to fly, it would become "the NGST mission" and the larger telescope would never be built.

With Nexus cancelled, the project turned to a substantial ground demonstration program using the wavefront sensing and control system described in Sect.17.

The Nexus concept was a full-fledged observatory and was later adapted to an Earth science mission proposal sited in geostationary orbit, but it was never implemented[72].

## 16. The Pre-Phase A architectures

On July 7, 1998, NASA selected TRW and Ball Aerospace to conduct two independent studies of the NGST mission but they eventually teamed up. To stay in the competition, Lockheed Martin did a parallel study on their own funds.

For its study, Lockheed Martin adopted a primary mirror configuration and deployment similar to those of the NGST "Yardstick", with petals folded up and down for launch. In this case, though, the secondary mirror was "fixed", as it fitted inside the fairing of an Atlas V launcher[73] (Fig. 31). The sunshield was composed of 6 membrane layers in a diverging V geometry to allow entrapped heat to migrate out. It was deployed by unrolling mechanisms.

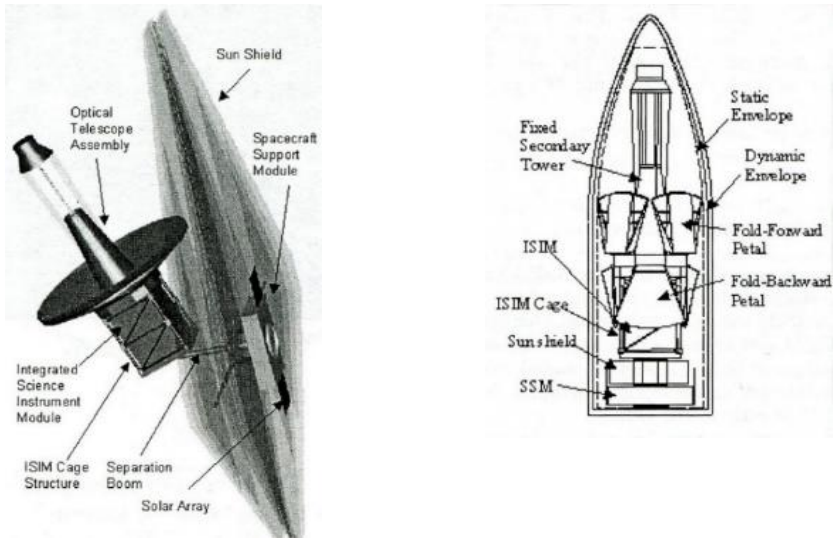

**Fig. 31.** The Lockheed Martin Phase A concept, deployed at left and stowed for launch at right.

In the TRW/Ball architecture, the primary mirror was composed of hexagonal segments divided into three sections. These were stowed for launch by folding in a drop-leaf or chord-fold table fashion (Fig. 32)[74]. The sunshield was composed of 5 layers which unfurled after launch.

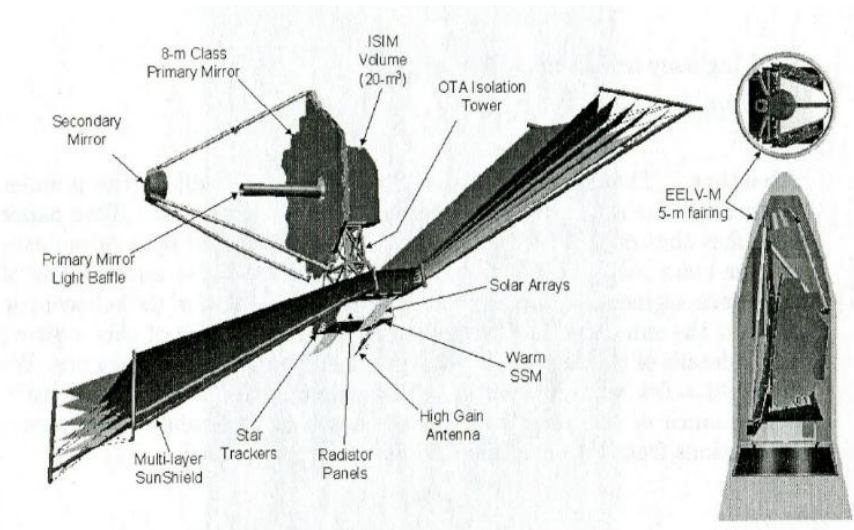

Fig. 32. The TRW/Ball Phase A concept, deployed at left and stowed for launch at right.

## 17. Architecture and technology hand in hand

Although industry feasibility studies had concluded that no new inventions were required to implement the proposed NGST architecture, it was clear that many areas would need extensive development and validation, as is almost always the case for new space missions. Several concerned the science instrument package (e.g. cryo-cooling for mid-IR detectors; low noise, large format detectors; microshutters) and will not be covered here. We highlight a few others where significant technological advances were made for JWST and which could be useful in future space missions.

- *Beryllium primary mirror segments*

Beryllium was chosen for all mirror substrates because of its exceptional structural and thermal qualities (see Sect. 18). It was not the first time it had been used for telescope mirrors. The Infrared Astronomical Satellite (IRAS) launched in 1983 had a 60-cm beryllium mirror. Subsequently, the 1.1-meter secondary mirrors of the four Very Large Telescopes (VLT) of the European Southern Observatory built in the late 1990's were also made of lightweighted beryllium[75] . This was to save weight compared to glass but, more importantly, to lower inertia, as IR observations from the ground require the secondary mirror to be rocked at 10-50 Hz for "target-sky chopping". The VLT secondary mirrors had been successfully fabricated and polished, but their surface had to be nickel-plated first, as regular grade beryllium has a rough surface which cannot be polished to optical quality. That technique could not be used for JWST since the beryllium/nickel thermal expansion mismatch would have led to unacceptable figure distortion at cryogenic temperatures.

Given that large lightweighted beryllium mirrors operating at cryogenic temperatures are necessary components of a number of space missions, NASA and the US Department of Defense initiated a technology development program in 1998 to validate this key technology. This involved developing a new grade of beryllium, O-30, with a much finer microstructure permitting excellent polish. The O-30 grade beryllium had the additional benefit of being more isotropic than its S-200F predecessor, making its behavior much more predictable and amenable to accurate modeling. The validation of the fabrication and polishing methods for this new material was accomplished through a series of small and full-scale demonstrators and completed in 2004[76,77,78].

Nonetheless, while this was a crucial development, the actual production of the beryllium mirror blanks and their polishing and cryogenic figure measurement was a major project, and it is to NASA's credit that it initiated that activity in Phase B, earlier than normal.

- *Cryogenic actuators for mirror control*

The primary mirror segments needed to be positioned in all six degrees of freedom and adjusted for curvature so as to conform to the prescribed overall mirror figure. This required actuators capable of 10 nanometer position resolution over a range of 20 mm and operating under cryogenic conditions (30K). Such actuators did not exist and a device was developed incorporating traditional components such as stepper motors, gears, bearings and flexures, but in a unique configuration which is covered by two patents[79]. The remarkable actuators were submitted to multiple pre-launch tests, both at room and cryogenic temperatures[80,81], and they deployed and performed perfectly on obit.

- *Primary mirror backplane support structure*

JWST has no adaptive optics system to correct for wave front errors due to attitude changes (see Sect. 18). This means that the structure supporting the primary mirror segments must be extremely stable, as its deformation results in direct degradation of the mirror figure. In order to avoid too frequent readjustments of the optical system, deformation of the backplane needs to be less than 25 nm over the telescope's full slew range, 3 orders of magnitude more stable than for typical telescopes. The backplane is made of carbon fiber composite, a material chosen for its strength, stiffness, and light weight. The use of composite for telescope structures was not new (it had been used in HST), but designing them to be super-stable and to survive the very large stresses that build up during on orbit cooldown (from 20 C to 40 K) was new. A new way to measure stability was also needed, so the structure could be characterized independently of the mirrors. Special analysis and measuring tools were developed and a 1/6th size cut-out of the primary mirror structure was built to demonstrate that the stability requirement was met[82,83,84].

- *Wavefront sensing and control*

For a segmented mirror to function as a single mirror surface with the proper optical figure, the segments need to be co-aligned, co-focused and co-phased[85]. Achieving this is routine on large ground segmented mirror telescopes. However, they benefit from special measuring equipment and possible human intervention in anomalous cases whereas JWST had to do it on its own, relying solely on science instruments data. A specific process called Wavefront Sensing and Control (WFSC) was created for that purpose by teams from JPL, GSFC, and Ball Aerospace, relying on dispersed-fringe spectral methods to phase the primary segments, and image-based phase retrieval techniques for high-resolution wavefront sensing and control. The latter were derived from methods developed in 1992 to measure the spherical aberration that HST suffered at launch[86]. This complex system is described in several papers[87,88,89]. It was first validated using surrogate image data, then a subscale 3-segment testbed was built for a hardware demonstration[90]. Finally, a full-scale, on-sky demonstration was performed using the Keck Telescope, and a larger 18-segment testbed was built during the final design phase to further refine the process[91,92]. The WFSC system, which was a critical component of JWST's architecture, has worked on orbit with remarkable success and the methodology is now thoroughly validated[93].

- *Large deployable multilayer sunshield*

The very large sunshield of JWST, made of ultra-thin layers, extremely cold on the telescope side (40K) and burning hot on the sun side (360K), was a critical component of the mission. If the deployed sunshield layers did not have the desired shape and spacing, the thermal performance would have been seriously degraded. If the shield did not fully deploy or if layers should touch each other or tear, the mission would be lost. It was a major engineering challenge because of the tight dimensional constraints on a flimsy, multi-layered, very large deployable element. The sunshield could be analyzed by structural and thermal modeling and its subsystems could be tested independently, but it was too large to be tested as a whole in a vacuum chamber. The solution was to test a series of engineering models at room temperature and a 1/3 scale model that could fit inside a vacuum chamber[94]. After launch, the complex system with its 8 motors, 90 cables and 400 pulleys deployed flawlessly. A remarkable accomplishment.

## 18. Changes to the architecture during final design

In December 2001 the program formally began the transition from feasibility studies to the design phase. The TRW/Ball architecture was selected, with a 7-meter diameter primary mirror composed of 36 segments in 3 rings. In September 2002 the prime contract for construction of the observatory was awarded to Northrop Grumman, which had acquired TRW the previous July, and the project name was changed to JWST. (By amusing coincidence, at roughly the same time, the Northrop Grumman division in charge of the project changed its name to Northrop Grumman Space Technology, aka NGST.)

By early 2003 it had become clear that a descope was unavoidable if the budget was to be met and it was eventually decided to reduce the collecting area of the primary mirror from about 30 down to 25 square meters. One solution was to use an oval shaped segmented primary mirror that did not deploy. But the preferred solution was to reduce the number of segments from 36 in 3 rings to 18 in 2 rings, with an overall aperture of 6.6 meters in diameter[95]. Reducing the number of segments had the advantage of having a smaller edge perimeter, and edges were one the most difficult parts of a mirror to make. This reduction led to hexagonal segments of 1.3-meter flat-to-flat, the maximum dimension which could be accommodated by the autoclave used for fabrication of the mirror blanks should they be made of beryllium.

A related consideration was that the baseline primary mirror segments had only 3 degrees of freedom: piston, tip and tilt. This was risky because it imposed constraints on the mirror segments' optical figure. It was therefore decided to use hexapod mounts providing 6 degrees of freedom, their extra cost being compensated for by the smaller collecting area and reduced number of segments.

That smaller collecting area was considered compatible with the scientific goals of the observatory, but it could not shrink any further. Table 2 summarizes the evolution of the primary mirror size of the principal HST successor candidates over the years, together with the other main observatory parameters concerning spectral coverage and sensitivity. Values for HST are included for comparison.

**Table 2.** Sensitivity and spectral coverage goals for HST, NGST and JWST, expressed in design parameters.

|  | HST | NGST | | | JWST | |
| --- | --- | --- | --- | --- | --- | --- |
|  |  | 1985 | 1996 | 2001 | 2003 | 2005 |
| Primary mirror diameter (m) | 2,4 | 10 | 8 | 7 | 6.6 | 6.6 |
| Wavelength of diffraction limit (µm) | 0.5 | 0.12 | 2 | 2 | 1 | 2 |
| Spectral coverage (µm) | 0.12 to 2.5 | 0.12 to 10 | 0.6 to 28 | 0.6 to 28 | 0.6 to 28 | 0.6 to 28 |
| Optics temperature (K) | 294 | 130 | 30 | 30 | 40 | 40 |

The entire architecture of JWST was revisited during the design phase and many studies and trades were performed[3,96].

One early study assessed the choice for the substrate material for the primary mirror segments. Detailed tests were conducted to compare beryllium to ultra low-expansion glass from the point of view of performance, manufacturing time and cost. That study confirmed beryllium as the best option because of its lighter weight and superior properties at cryogenic temperatures (better thermal conductivity and smaller coefficient of thermal expansion)[97,98].

Another study dealt with the stability of the optics. As the telescope points to a new target its attitude with respect to the sun changes, resulting in temperature variations in the optics and supporting structure and, potentially, image quality degradation. To correct for this, JWST has both active optics (with a fine-steering mirror) and built-in corrective optics (with its primary mirror segments positioning and shaping actuators). It would also have been possible to introduce a deformable mirror in the optical train for adaptive wavefront control, as had been proposed for the 10-meter telescope concept of Sect. 3 and for the NGST Yardstick concept. The use of adaptive optics features at cryogenic temperatures was judged too complex, however, and the observatory was instead designed for passive structural-thermal stability.

Perhaps one of the most difficult trade studies involved placement of the processing and control electronics for the science instruments. The majority of these were flight qualified at room temperature, and thus were initially placed on the "hot" side of the observatory (Fig. 33). However, concerns quickly arose over the electrical noise that could be incurred from cable lengths on the order of 10 meters. Also, since the observatory would have to be tested in two parts, a cryogenic portion and a room temperature portion, the final performance testing of the science instruments with all of their 1500 plus electrical connections in final flight configuration would not be possible. This electrical problem was solved by placing the critical science instrument processing and control electronics in a highly insulated heated compartment on the cold side of the observatory. The design of the additional insulation for the instrument electronics was the major challenge in the thermal design of JWST's architecture.

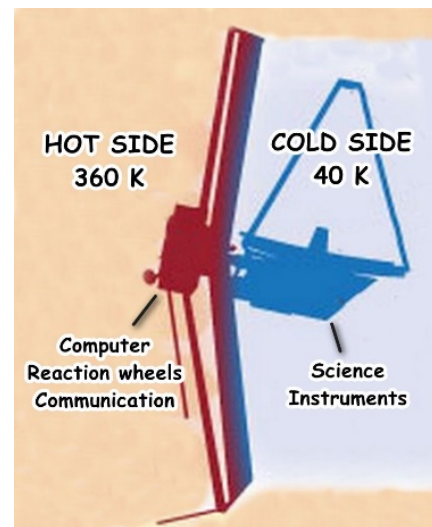

**Fig. 33.** Hot and cold sides of JWST.

In addition to this thermal problem, the overall low cooling margins of the instrument radiators were a concern. Parasitic heat loads are notoriously hard to predict in cryogenic instruments, and by the Critical Design Review of the observatory in April of 2010, their estimated values had grown dramatically, degrading the radiator margins to unacceptably low levels[99]. This was solved by replacing the instrument radiators facing the sunshield with a deployable radiator facing deep space. This last architectural change restored radiator margins and decoupled their performance from the sunshield.

Aside from several changes in the Science Instruments package which are not covered here, the most important modification of the original Phase A architecture concerned the sunshield.

When slewing from one target to the next, the observatory uses flywheels (also called reaction wheels) which are accelerated or decelerated to transfer angular momentum from the wheels to the observatory. Momentum changes can be partially managed by observing at an orientation that builds momentum in a particular direction, followed by an observation at an orientation that counters that momentum. But after a while, the wheels' spin speeds can reach their maximum limit and they need to be "desaturated". On HST this was done by creating a counter torque with electromagnetic rods working against the Earth's magnetic field. But far from Earth, such as at L2, that is not possible, so the reaction wheels on JWST are desaturated by firing thrusters. And since propellant is a critical resource on JWST, it was decided to reduce the amplitude of the momentum required for attitude changes by downsizing the sunshield by about 30% in area and adopting a 3-plane configuration in lieu of a single flat one. As an additional benefit, this reduced the overall mass of the observatory.

The reduction in shield size had an observational impact, however, because although the sunshield would still fully protect the telescope from the Sun, there are times when the Earth and the Moon could shine on the telescope (Fig. 34), increasing the background to above zodiacal light, in particular in the infrared. To avoid this situation, the "field of regard" (the range of orientations that the observatory could point) was reduced, although still enabling 50% sky coverage at any time and 100% sky coverage over a year.

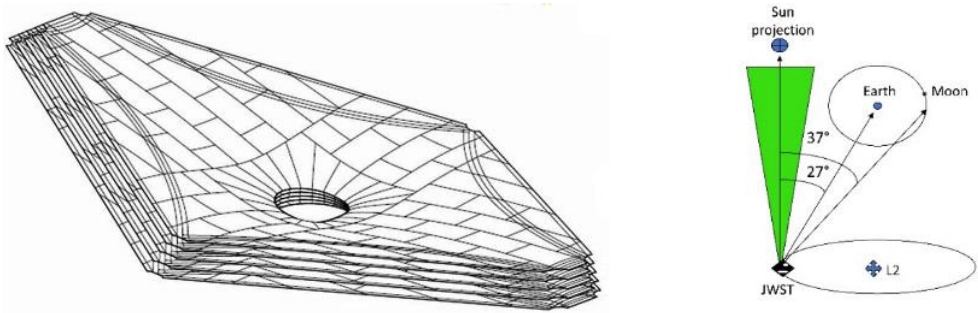

**Fig. 34** At left, the new shield seen from the telescope side. The external layer is relatively flat, the inner one is more curved. At right, worst case angular positions of the Earth and Moon when not hidden by the sunshield.

Another modification of the sunshield was the addition of a trim or momentum flap. As solar photons strike the large, highly reflective sun shield, they exert a pressure on the observatory as a whole (Fig. 35).

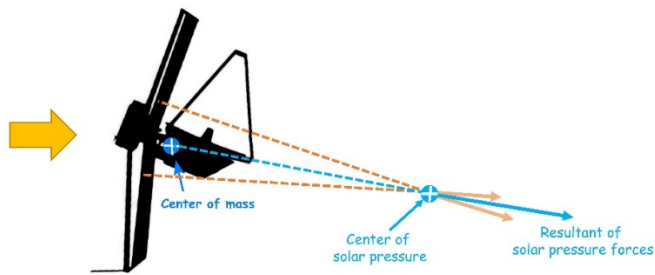

Fig. 35. Solar radiation pressure exerts a force which is essentially perpendicular to the surfaces it impinges on. For the observatory to be balanced, the resultant force acting on the sunshield and the space support system must pass through the center of mass.

This pressure pushes the observatory out of its orbit, an effect that is corrected during station keeping maneuvers. But it also creates a torque when the resultant force of the solar radiation pressure on the sunshield and equipment on the hot side does not pass through the observatory's center of mass.

The attitude control system counteracts this torque by changing the spin speed of the reaction wheels but, just as in the case of observational slews, this comes at the cost of propellant consumption. To minimize solar radiation torque, it was first envisaged to continuously balance solar pressure by adjusting the pitch of the front part of the sunshield, but this proved to be too complex, risky and had a major impact on the thermal stability of the telescope. Instead, the shape of the sunshield was modified to reduce overall solar pressure and a deployable flap was added to the aft of the sunshield. The pressure exerted on the large reflective momentum flap reduces the offset, as explained in Fig. 36. This flap is not movable on orbit, but its post-deployment angle was adjusted just prior to launch based on final measured mass properties. The compensation is not exact, but it minimizes the momentum build-up over the range of pointing between scheduled momentum dumps.

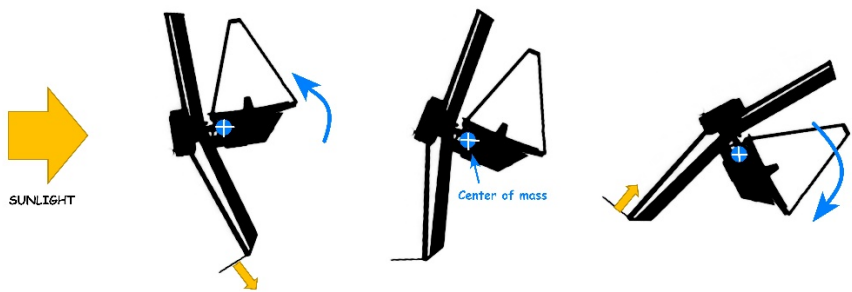

**Fig. 36**. The resultant of the solar pressure on the momentum flap (yellow arrow) creates a torque (blue arrow) that compensates the imbalance of solar radiation pressure when the observatory changes attitude. The flap acts like the rudder of a sailboat after the fore and aft sails have been trimmed, compensating the residual misalignment of the wind effort and the drag of the hull along the direction of travel.

After all these design changes, JWST had become smaller than in the original Phase A version while superficially retaining its original look (Fig. 37). But many of the changes had been significant.

In addition, we must add that the changes here reported were not the end of the story. The final design involved addressing many devilish details that required hard work and creativity, arguably more so than for other space observatories because of the complexity and novelty of the design[99].

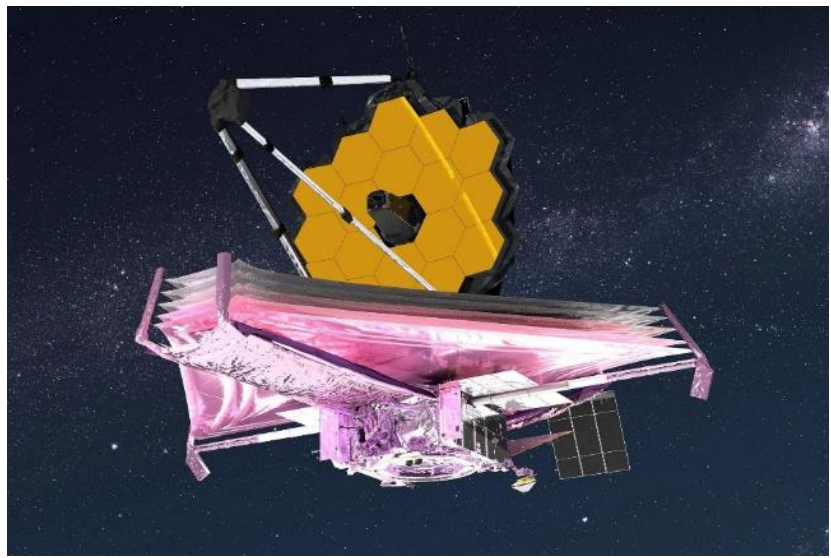

**Fig. 37** Artist's view of JWST deployed on orbit showing the space support unit, the sunshield booms, and the trim flap. (NASA)

## 19.  The real thing, on orbit

The detailed design, construction and testing of JWST took almost 20 years, and it was finally launched on December 25, 2021.   Deployment, orbit transit and commissioning went largely according to plan, and temperature, image quality, and straylight levels are better than expected[3,96].  In particular, thanks to the wavefront control system, excellent performance of the mirror actuators, effective jitter suppression, unprecedented pointing stability and passive thermal stability, the telescope is diffraction-limited to 1 μm – surpassing the original design goal of 2 μm.

And thanks to an efficient sunshield, the inner layer of which is at around 55 K, the temperature of the optics ranges from 35 to 55 K, allowing broad-band observations to be background limited by zodiacal light up to 12.5 μm wavelength[100,101].

Moreover, careful design and construction of the entire observatory has resulted in extremely stable optics.  In spite of variable thermal states and dynamic excitations, the stability of the primary mirror segments is superb[102], requiring realignment only every month or two.

A favorable launch date, perfect Ariane 5 launch and error-free commissioning has given JWST an abundance of propellant reserve.  Combined with the regular usage that is lower than planned, the observatory will have propellant for more than 20 years of active life.

One consequence of the open architecture is that JWST is exposed to micrometeoroids that can damage the mirrors. This was anticipated and the resulting degradation had been considered acceptable. Additionally, with its wavefront control system, JWST has the ability to detect and correct, to some extent, mirror deformations caused by impacts.  But a major micrometeoroid strike in May 2022 was sufficiently worrisome, especially in view of JWST's new longer lifetime, to begin avoiding observations in the sector of the orbital direction of motion where the velocity of JWST adds to that of micrometeoroids, potentially quadrupling the energy of the impact. Fortunately, the frequency of large strikes has remained small, becoming more consistent with pre-launch calculations.

Artifacts due to light scattered from bright sources have occasionally been seen in the near-IR camera but they are due to an uncoated optical element in the instrument, not to the open architecture, where the optical system is well baffled.

The observatory is now in the hands of the astronomy community and most observations fall within categories for which the observatory was optimized (Sect. 11).  Some were unanticipated, however, such as those for exoplanet science and detailed galactic studies.   And these are well met by the observatory as designed, proof once again that any observatory which opens a new window on the Universe, such as JWST with its IR coverage at high sensitivity, is bound to lead to discoveries in domains for which it was not originally intended.

The mid-IR instrument is used for about 25% of the observations, fully justifying its inclusion and protection from any descope.   This is especially important since the mid-infrared, like the near-infrared, is beyond the reach of any other facility, ground or space, at the sensitivity afforded by JWST.

## 20. Conclusion

As has been shown, when the go-ahead for the project was finally given the architecture of JWST did not suddenly burst into bloom, fully formed, like a flower in spring.  Over the course of almost 20 years before JWST entered Phase-B in 2003, no fewer than 16 unique independent mission concepts were studied.

And subsequently, over nearly two decades, the industry modified and improved the basic architecture as they turned a concept on paper into a living, breathing machine.  Two concepts were studied in detail during Phase A, and the final flight implementation was the result of at least 4 major redesigns and multiple small revisions.  In the end, thanks to experience, creativity, know-how, and hard work, a machine was created that now accomplishes the scientific goals originally set by the astronomical community.

The initial 8- to 10-meter target may have been too ambitious but, as NASA's Administrator Dan Goldin told some of us in 1999: "My role is to be a locomotive and pull you into new territory. If your train car can't keep up, so be it - I'll slow down. But we will all have gone as far as we possibly could."

So true! Although not as large as we had initially aimed for, look what we do have now: a novel architecture that can be expanded to ever larger observatories of the future. JWST is not simply the "next generation space telescope", as it was once called. It is a whole new breed of telescope.

And what a miraculous success it is – with its smaller brother HST still up there to boot. Giacconi would surely be pleased !

## Acknowledgments

What has been reported here is the story of the JWST morphology, mostly from the point of view of the technical people charged with formulating its architecture. They toiled at their "drawing boards" but were constantly helped and inspired by managers, scientific oversight committees, scientists and colleagues at their institutions, people they met at conferences, and by the experience gained by their predecessors who worked on the Great Observatories, HST, Chandra and SIRTF.

So, over the years, a great many people contributed to the formulation of the JWST architecture. We wish to particularly acknowledge Roger Angel, John Bolton, Christopher Burrows, Richard Capps, John Campbell, Daniel Coulter, Rodger Doxsey, Herbert Gursky, David Jacobson, Michael Krim, Paul Lightsey, John MacKenty, Scott May, Aden Meinel, Gary Mosier, Matt Mountain, Charles Perrygo, François Roddier, Phillip Tulkoff, Ray Wilson, and Robert Woodruff for many fruitful discussions during the conception phase.

We are grateful to Matt Mountain and Robert Smith for their helpful comments and to Sally Bely for editing the manuscript.

---

**BIOGRAPHY OF THE AUTHORS**

**Pierre Y. Bely** is a French-American engineer trained at the Ecole Centrale de Paris and holding an MS in engineering from UC Berkeley. After serving as Project Engineer for the design and construction of the Canada-France-Hawaii Telescope in Hawaii, he joined STScI in 1984 as Chief Engineer in charge of the technical monitoring of HST and, in parallel, worked on the earliest conceptual studies for NGST. From 1996 and until his retirement in 2000 he held the position of Mission Architect for NGST. He is the main author of one of the few textbooks on large optical telescope design and construction.

**Jonathan Arenberg** is currently the Chief Mission Architect for Science and Robotic Exploration at Northrop Grumman. Prior to working on JWST, he worked on the Chandra X-ray Observatory and co-invented the starshade. During his tenure on JWST he held several positions, Design Integration Lead, Observatory Systems Engineering (OSE) Deputy, OSE Manager, ultimately becoming Chief Engineer. He is an Associate Fellow of the AIAA and a Fellow of SPIE.

**Charles (Charlie) Atkinson** spent 24 years on JWST, most recently as the JWST Chief Engineer after being the Deputy Telescope Manager. Before JWST, Charlie was responsible for the integration and alignment of the Chandra X-Ray Telescope, among other EO systems. Charlie has been awarded the Robert H. Goddard Exceptional Achievement Award in Engineering, the NASA Exceptional Public Service Medal, the AIAA Goddard Astronautics Award, the Aviation Week Program Excellence Award and the NASA Distinguished Public Service Award.

**Richard Burg** is a physicist with a Ph.D. from the Massachusetts Institute of Technology. He was a Postdoctoral Fellow with the Space Telescope Science Institute working for Riccardo Giacconi on spectroscopic classification of galaxies, the X-ray background, and X-ray optics, then went on to work on the architecture and technology development of NGST at NASA GSFC.

**Mark Clampin** holds a Ph.D. in astronomy from the University of Saint Andrews in Scotland. After serving as HST instrument scientist at STScI, he joined NASA/GSFC as JWST Observatory Project Scientist, and subsequently rose to the position of Director of the Astrophysics Science Division. He is currently the Director of the Astrophysics Division at NASA Headquarters in Washington, D.C.

**Lee Feinberg** holds an M.S. in Applied Physics from Johns Hopkins University. He joined NASA's Goddard Space Flight Center where he worked on HST's corrective optics and science instruments, then as the Optical Telescope Element Manager for the James Webb Space Telescope for more than 20 years. He is currently the Principal Architect for the future Habitable World Observatory (HWO).

**Paul Geithner** is an engineer with a BS from Virginia Tech and an MBA from the University of Virginia. He Joined NASA's Goddard Space Flight Center in 1991 where he worked on HST repair and upgrade missions, then as Chief Systems Engineer as part of the NGST team and subsequently on JWST, rising to the position of Deputy Project Manager – Technical.

**Garth Illingworth** is an astronomer with a Ph.D. in astrophysics from the Australian National University. After working at the Kitt Peak National Observatory, he served as Deputy Director of STScI from 1984 to 1987 and played a leading role in the early development of NGST. In 1988 he was awarded a professorship at the University of California at Santa Cruz but continued to be involved in the development of NGST/JWST and chaired the JWST Science Advisory Committee from 2009 to 2017.

**John Mather** is an astrophysicist with a Ph.D. in Physics from the University of California in Berkeley. He joined NASA's Goddard Institute for Space Studies at Columbia University in New York City and worked on the conception, design, implementation and data analysis of the Cosmic Background Explorer (COBE) satellite to map the cosmic microwave background (a work for which he was awarded the Nobel prize, together with George Smoot). From 1995 to date, he has been Project Scientist for NGST/JWST at GSFC, supervising the scientific interests of the project and leading the Science Team of the mission.

**Michael Menzel** is an engineer with an MS in Physics from Columbia university. He has held a variety of positions in the aerospace industry, in particular at Lockheed Martin as Deputy Program Manager for the HST Servicing Missions and Chief Systems Engineer for their Pre-Phase A and Phase A studies of NGST, and at Northrop Grumman as member of the JWST Systems Engineering Team. He joined NASA's GSFC in 2004 as systems engineer for JWST overseeing the architecture, design, integration and systems validation of the mission.

**Max Nein** holds an MS degree in mechanical engineering from the Technical University of Munich. He joined the NASA Marshall Space Flight Center in1962 during a transfer from the US Army Ballistic Missile Agency (ABMA). At ABMA and at MSFC he worked on cryo-propellant flow and heat transfer problems of the Apollo lunch vehicles, Skylab, Solar Telescopes on ATM, the Hubble Space Telescope design and other space flight systems until his retirement from NASA in 1999. Subsequently he worked on NGST under contract to industry until 2006.

**Larry Petro** is an astronomer with a Ph.D. in Astronomy from the University of Michigan. After holding the position of research scientist for the SAS-3 X-ray satellite, he joined STScI in 1986 as a member of the HST astronomy staff. He made important contributions to the NGST conceptual phase and served as STScI mission scientist for JWST from 1996 until 2006. From 2006 to 2012 he was an instrument scientist for HST's WFC3. He then served as a program officer at NASA HQ Astrophysics Division until retiring in 2015.

**David Redding**, who holds a PhD in Aeronautics and Astronautics from Stanford University, joined NASA's Jet Propulsion Laboratory in 1992 and has extensive experience in active optics. His work on wavefront sensing and control systems and integrated modeling was crucial to JWST and has significantly contributed to advancements in space telescope technology.

**Bernard Seery,** formerly of NASA Goddard Space Flight Center, is currently Vice President for Science and Technology, Universities Space Research Association.

**Philip Stahl,** who holds a PhD in optical science from the University of Arizona, is a Senior Optical Physicist at NASA Marshall Space Flight Center. He is a leading authority on optical metrology, optical engineering and phase-measuring interferometry, and was responsible for the development of mirror technologies for JWST.

**Massimo Stiavelli** is an astronomer with a Ph.D. in Physics from the Scuola Normale Superiore in Pisa, Italy. In 1985 he joined STScI as a member of the HST astronomy staff and was the main contributor to the formulation and analysis of the NGST Design Reference Mission. Between 2003 and 2012 he held the position of Project Scientist for JWST and was the JWST Mission Head at STScI from 2012 to 2024.

**Hervey S. (Peter) Stockman** is an astronomer with a Ph.D. in Physics from Columbia University. After working on the design of instruments for ground-based telescopes at the University of Arizona he joined the astronomy staff of STScI in 1984 and was Deputy Director of the Institute from 1988 to 1995. He was the prime mover behind the High Z mission proposal and led the development of the NGST concept. He later served as a member of the JWST Scientific Oversight Committee until his retirement.

**Scott Willoughby** is an engineer with an MS in Communication Systems from the University of Southern California. In 1989 he joined TRW where he held the position of project manager for a variety of space missions. After TRW merged with Northrop Grumman, he became program manager for JWST in 2009 and was later named director of the program. His role was pivotal in overseeing the final phase of the design of JWST and its construction. He was elected to the National Academy of Engineering for his efforts on JWST.